\documentclass[a4paper,10pt]{article}
\pdfoutput=1 

\usepackage{jheppub}
\usepackage{subfigure}
\usepackage[T1]{fontenc}

\title{Holographic mutual information in global Vaidya-BTZ spacetime}

\author{Vaios Ziogas}

\affiliation{Centre for Particle Theory \& Department of Mathematical Sciences,

Durham University, South Road, Durham DH1 3LE, United Kingdom}

\emailAdd{vaios.ziogas@durham.ac.uk}

\abstract{We investigate the evolution of the mutual information between two spatial subsystems in a compact 1+1 dimensional CFT after a quantum quench. To this end, we use the dual holographic process, given by the 2+1 dimensional Vaidya-BTZ spacetime in global coordinates, which describes the collapse of a spherically symmetric null shell. So, we first discuss the spacelike geodesic structure of this geometry and then we present the various behaviors of the holographic mutual information observed in this case. We also consider the analogous process in the adiabatic limit and compare these two cases from a geometrical point of view.}


\preprint{DCPT-15/45}

\begin{document}
\maketitle
\flushbottom

\section{Introduction}\label{sec:intro}

There has always been a wide interest in non equilibrium phenomena in quantum field theory, since they can be used to describe processes in various areas of physics, including condensed matter and high energy physics. In particular, it is very important to gain a deeper understanding of thermalization, meaning the way in which physical quantities attain their equilibrium values after some kind of perturbation has acted upon an initial equilibrium state. Of course, QFT implies that a pure state will remain pure under time evolution, so it is interesting to understand precisely how the final pure state will resemble a thermal one.

A simple model of a thermalization process is the so-called ``quantum quench'', which involves a change in the parameters of the Lagrangian of the system, either instantaneously or over a finite period of time. For example, it can describe a uniform or point-like energy injection, depending on the kind of sources that are turned on. So, the vacuum state of the initial Lagrangian will be a high energy state of the final Lagrangian, and it will evolve with the final Hamiltonian.

It would be desirable to study the above problems in the strong coupling regime of QFTs, where perturbation analysis becomes inapplicable. Although some progress had been made using QFT techniques, this problem became more tractable with the appearance of the proposal that certain strongly coupled $d$-dimensional quantum field theories with a large number of degrees of freedom are dual to classical gravity theories in $d+1$ dimensions (this idea originated in \cite{maldacena}--\cite{gubserklebanovpolyakov}; see \cite{hubenyads} for a recent review). Specifically, this so-called ``AdS/CFT correspondence'' suggests that the vacuum state of a conformal field theory in $d$ dimensions is dual to pure AdS$_{d+1}$, while a thermal state is dual to an (asymptotically AdS$_{d+1}$) black hole geometry. So, it makes sense that the process of gravitational collapse ending in a black hole in an asymptotically AdS$_{d+1}$ spacetime geometry is dual to thermalization in a CFT.

Apart from various correlation functions or physical observables, a particularly interesting object to study is the entanglement entropy. Even though it is not a physical observable and thus not directly measurable, it encodes the way quantum information is distributed in the system, as well as the quantum correlations between the subsystems. It is defined as the von Neumann entropy $S_\mathcal{I}=-\textrm{tr}(\rho_\mathcal{I}\ln \rho_\mathcal{I})$ of the reduced density matrix $\rho_\mathcal{I}$ corresponding to a subsystem $\mathcal{I}$. Usually we take $\mathcal{I}$ to be a spatial subregion of our theory and then $S_I$ measures the correlations of $\mathcal{I}$ with the rest of the system. $S_\mathcal{I}$ is typically divergent but, after regularization (usually by introducing a UV cut-off), we can isolate the finite part (this was investigated in \cite{srednicki}; see \cite{ecp} for a review). One can also define the mutual information of two regions $\mathcal{I}_1$ and $\mathcal{I}_2$ by $I(\mathcal{I}_1,\mathcal{I}_2)=S_{\mathcal{I}_1}+S_{\mathcal{I}_2}-S_{\mathcal{I}_1\cup \mathcal{I}_2}$. This object is interesting in its own right because it quantifies the correlations shared between $\mathcal{I}_1$ and $\mathcal{I}_2$ but not the rest of the system, it is an upper bound for bounded correlators between $\mathcal{I}_1$ and $\mathcal{I}_2$, and it is finite as long as the two subsystems are not adjacent.

In general, the entanglement entropy is very hard to compute with QFT techniques, and it has only been done for special theories (e.g. free scalar field theories, CFTs), in special states (e.g. vacuum or thermal) or for specific regions with some degree of symmetry (see for example \cite{calabresecardy2}, \cite{casinihuertamyers} among others). However, it was realized that holography greatly simplifies this problem.

For generic theories with non-static Einstein-Hilbert gravity duals, the holographic entanglement entropy (HEE) $S_\mathcal{I}$ of a boundary region $\mathcal{I}$ (not necessarily on a constant $t$-slice) is computed using the ``HRT'' proposal of \cite{hubenyrangamanitakayanagi}. According to this proposal, to order $\mathcal{O}(G_N^{-1})$, the HEE is given by the area (in Planck units) of an extremal surface $\mathcal{A}$ anchored on the boundary $\partial\mathcal{I}$ of $\mathcal{I}$ (i.e. with $\partial\mathcal{A}=\partial\mathcal{I}$) and homologous to $\mathcal{I}$. Specifically, the homology constraint requires the existence of an everywhere spacelike manifold with boundary $\mathcal{A}\cup\mathcal{I}$\footnote{See \cite{hubenymaxfieldrangamanitonni} for a discussion of some subtleties (which imply that this condition requires some refinement).}. In case there are many surfaces satisfying these conditions, we should choose the minimal one, so we can write $S_\mathcal{I}=\min_\mathcal{A}\frac{\textit{area}(\mathcal{A})}{4G_N}$. It is easy to see that if the spacetime is static, the HRT proposal coincides with the ``RT'' proposal of \cite{ryutakayanagi1} and \cite{ryutakayanagi2}, replacing the extremal surface by a minimal one on a Euclidean constant-time bulk slice. Note that the choice of quantum state in which we compute the entanglement entropy is implicit in the choice of a bulk gravity solution, and that the regularization procedure analogous to introducing a UV cut-off in field theory is the introduction of a large radial cut-off $r_\infty$ in the bulk spacetime.

In this paper, we are going to investigate the evolution of the holographic mutual information (HMI) between two regions in a compact 1+1 dimensional CFT after a kind of quantum quench describing an instantaneous, uniform energy injection into the ground state of the system. We are going to use the simplest possible dual gravitational analogue: the collapse of a spherically symmetric, infinitely thin null shell in pure 2+1 dimensional AdS spacetime, resulting in the creation of a BTZ black hole. We will ignore the details of the matter which gives rise to the so-called global Vaidya-BTZ metric\footnote{Another model for holographic quenches was proposed in \cite{hartmanmaldacena}, but it will not be considered here.} describing the above spacetime, focusing only on the purely gravitational part, which is expected to give general and universal results. We will only consider this low dimensional model because it is easier to apply the HRT prescription in this case: the extremal surfaces are spacelike geodesics. However, it is able to capture important information about the higher dimensional cases too.

\begin{figure}
\centering
\includegraphics[width=.35\textwidth]{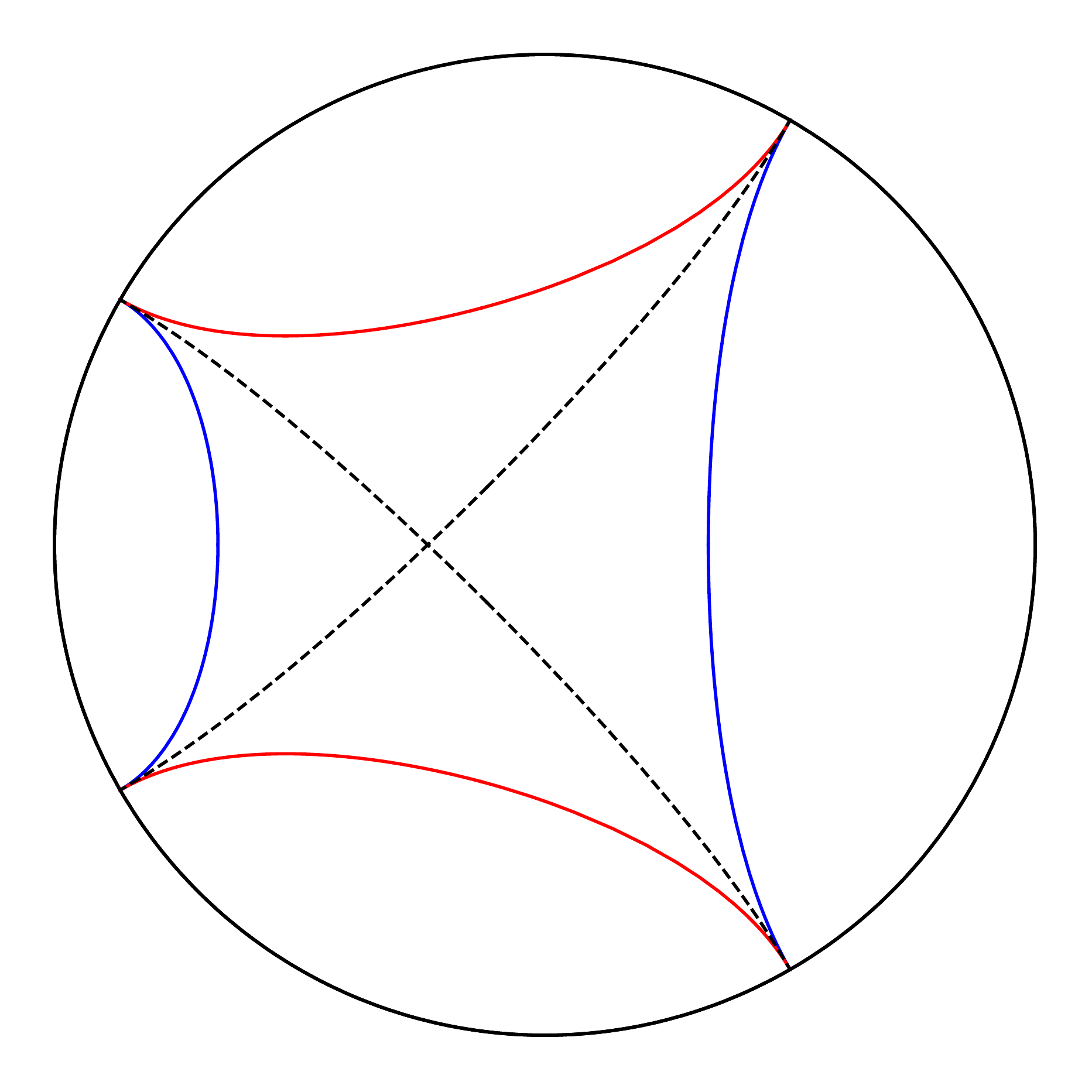}
\hfil
\includegraphics[width=.35\textwidth]{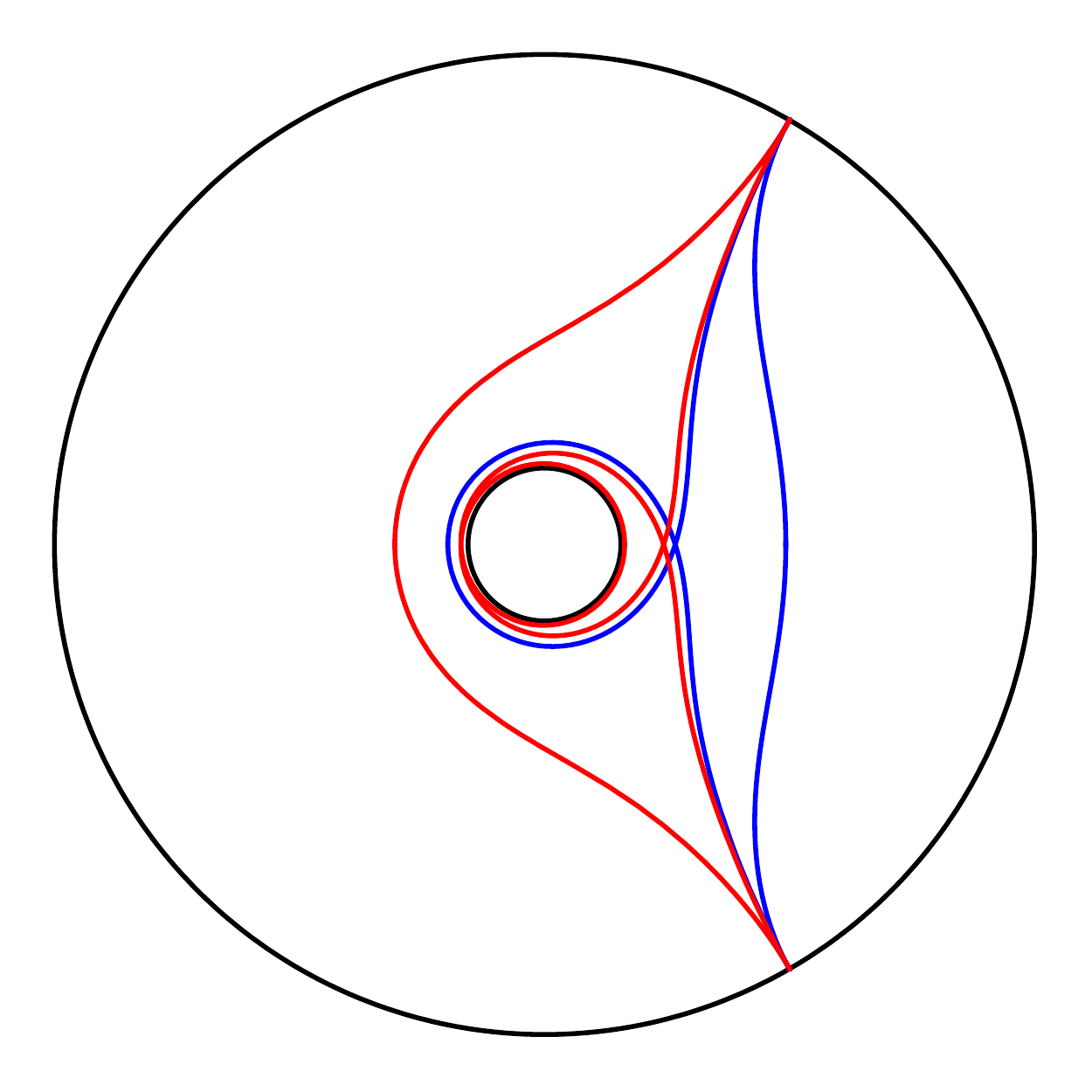}
\begin{picture}(0,0)
\setlength{\unitlength}{1cm}
\put(-0.1,2.5){$\mathcal{I}$}
\put(-7,2.5){$\mathcal{I}_1$}
\put(-5.6,2.5){$\bar{\mathcal{I}}$}
\put(-12.5,2.5){$\mathcal{I}_2$}
\put(-8.1,4.8){$A$}
\put(-8.2,0.2){$B$}
\put(-12.2,1.1){$C$}
\put(-12.2,4){$D$}
\end{picture}
\caption{\label{fig:1fourthbtz} Spatial projection of spacelike geodesics, with the boundary brought to finite distance using the coordinate $\rho=\tan r$. \textbf{Left:} Spacelike geodesics in pure AdS$_3$ spacetime, anchored on the endpoints $A$ and $B$ of the boundary interval $\mathcal{I}_1$ and $C$ and $D$ of $\mathcal{I}_2$ in various combinations. \textbf{Right:} Spacelike geodesics in BTZ spacetime with $r_+=1/4$. The inner black circle marks the event horizon of the black hole. The two blue geodesics wind around the black hole $0$ and $1$ times, while the two red ones wind around the black hole $0$ and $2$ times. The innermost blue geodesic is homologous to the smaller interval $\mathcal{I}$ and the outermost red geodesic is homologous to the complementary interval $\bar{\mathcal{I}}$.}
\end{figure}
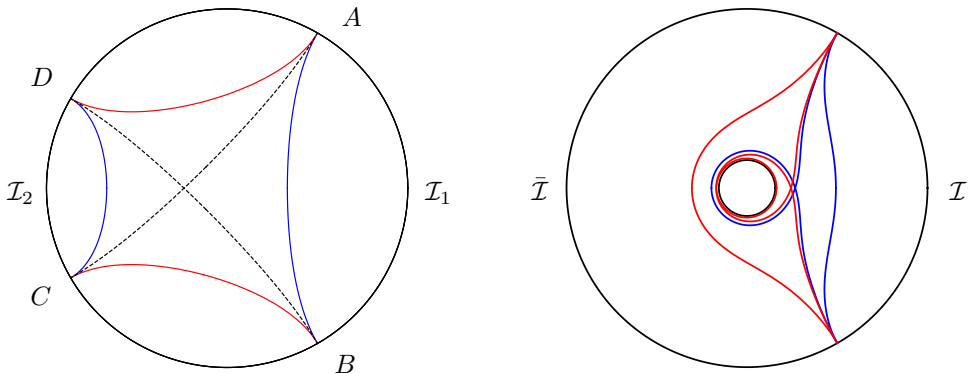

Some aspects of our general setup can be seen in figure \ref{fig:1fourthbtz} (left). A spatial slice of pure AdS$_3$ is presented, and two boundary intervals $\mathcal{I}_1$ and $\mathcal{I}_2$ have been singled out. Obviously, the HEE $S_{\mathcal{I}_1}$ of the interval $\mathcal{I}_1$ will be computed by the length of the spacelike geodesic $\left(AB\right)$ and $S_{\mathcal{I}_2}$ by the length of $\left(CD\right)$. However, for the union $\mathcal{I}_1\cup \mathcal{I}_2$ there are more candidates: both the set $\left(AB\right)$ and $\left(CD\right)$ (blue geodesics - ``connected'' configuration) and the set $\left(BC\right)$ and $\left(DA\right)$ (red geodesics - ``disconnected'' configuration) satisfy the boundary and homology conditions of HRT, so $S_{\mathcal{I}_1\cup \mathcal{I}_2}$ will be computed by the set with the smallest sum of lengths. Thus, we can write the HMI in the form
\begin{equation}\label{eq:formi}
\begin{aligned}
I(\mathcal{I}_1,\mathcal{I}_2) & =(AB)+(CD)-\min[(AB)+(CD),(BC)+(DA)]\\
& =\max[(AB)+(CD)-(BC)-(DA),0].
\end{aligned}
\end{equation}
We can immediately see that eq. \eqref{eq:formi} implies $I(\mathcal{I}_1,\mathcal{I}_2)\geq0$. It is important to note that the dashed geodesics $\left(AC\right)$ and $\left(BD\right)$ do not satisfy the homology condition (there is no smooth, everywhere spacelike interpolating manifold between the geodesics and $\mathcal{I}_1\cup \mathcal{I}_2$). Of course, the Vaidya-BTZ spacetime that we will consider is not static, so the geodesics will not lie on a constant time slice as in figure \ref{fig:1fourthbtz}. However, even in this more complicated case, we should keep in mind that the ``diagonal'' geodesics $\left(AC\right)$ and $\left(BD\right)$ will not satisfy the homology condition because the interpolating manifold cannot be everywhere spacelike. So, the HMI will be computed as in eq. \eqref{eq:formi} in the rest of the paper.

The motivation for the current work is twofold. Firstly, even though similar models have been investigated extensively in the literature (see for example \cite{abajoapariciolopez}--\cite{liusuh2}), the main focus was on non-compact CFTs, (i.e. dual to Vaidya-BTZ in Poincar\'{e} coordinates). Significantly less attention has been directed to compact CFTs, where finite size effects may well result in unexpected behaviors for the HMI, since the generalization to compact space is generically non trivial. For instance, it is well known that the HMI in non-compact holographic CFTs typically exhibits a ``bump'' (see \cite{bbccg}, \cite{allaistonni}), so it is interesting to investigate whether more bumps, or qualitatively different behaviors, can appear in the compact case. Secondly, there is a sharp contrast between this, typically non monotonic, evolution and the evolution in the adiabatic approximation of the quench. As will also be explained later, the adiabatic approximation treats the system as always being in a thermal state with the temperature slowly increasing. The dual gravitational description of this process is given by a BTZ black hole with its radius slowly increasing (since the radius $r_+$ is proportional to the temperature $T$, $T\sim r_+$). We intuitively understand that, in this case, the ``growth'' of the black hole radius will ``push back'' the boundary anchored spacelike geodesics (since they cannot penetrate the horizon of a static black hole, see \cite{hubeny}). Combined with the fact that longer geodesics reach deeper into the bulk, this suggests that non monotonic behaviors for the HMI cannot appear in the adiabatic approximation. For example, we can see that if the ``connected'' configuration of geodesics dominates (i.e. $I=0$) initially, it will remain so during the entire evolution, in contrast to the behaviors observed in \cite{bbccg} and \cite{allaistonni}. It is interesting to investigate this problem in order to understand why these differences occur from a geometrical point of view.

For convenience, we are going to set the AdS radius of curvature $L_{AdS}$ to $L_{AdS}=1$, and also ignore factors of $4G_N$ below, loosely identifying the length of spacelike geodesics with $S_\mathcal{I}$. So we should keep in mind that the HRT formula for the HEE is a good approximation when quantum and higher curvature corrections are suppressed, since it is an order $\mathcal{O}(G_N^{-1})$ result in classical Einstein-Hilbert spacetime (with $G_N$ small). In other words, recalling that the central charge $c$ of the boundary CFT is given by the ``Brown-Henneaux relation'' $c=\frac{3 L_{AdS}}{2 G_N}$, we will be working in order $\mathcal{O}(c)$, in the limit of large $c$. As usual, the speed of light $\mathrm{c}$ and Planck's constant $\hbar$ are set to $1$, $\mathrm{c}=\hbar=1$.

The plan of this paper is as follows. In section \ref{sec:geoinvbtz} we start by presenting the Vaidya-BTZ spacetime and its spacelike geodesic equations. Then, we present the solutions of these equations, also describing the geodesic structure of the spacetime in some detail. In section \ref{sec:heeandmi} we plot our results for the holographic entanglement entropy and the holographic mutual information, before turning to the comparison with the adiabatic approximation. In the final section we conclude and we discuss some open questions. The appendix contains explicit calculations not presented in the main body of the work.

\section{Geodesics in Vaidya-BTZ}\label{sec:geoinvbtz}
\subsection{Vaidya-BTZ spacetime in global coordinates}
As explained in the introduction, we will consider the $2+1$ dimensional, asymptotically AdS spacetime described by the Vaidya-BTZ metric:
\begin{equation}\label{eq:vaidyabtz}
ds^2=-f(v,r)dv^2+2dvdr+r^2d\varphi^2,
\end{equation}
where $r$ is the usual radial coordinate with range $[0,\infty)$, $\phi\in[0,2\pi)$ is the angular coordinate with period $2\pi$, $v\in(-\infty,+\infty)$ is the ``ingoing time'' and
\begin{equation}\label{eq:f}
f(v,r)\equiv r^2-m(v)=r^2+1-\theta(v)(r_+^2+1)=\left\{
                                  \begin{array}{ll}
                                    r^2+1 & \equiv f_i(r),~~ \hbox{v<0} \\
                                    r^2-r_+^2 & \equiv f_o(r),~~ \hbox{v>0}
                                  \end{array}
                                \right.
,
\end{equation}
where $\theta(v)$ is the Heaviside step function. It is clear that $f(v,r)$ interpolates between pure AdS$_3$ for $v<0$ and the BTZ black hole with an event horizon of radius $r_+$ for $v>0$, while $v=0$ describes an incoming null shell which collapses and forms a curvature singularity at $r=0$ for $v>0$. More generally, we could consider a shell with non-zero thickness, by replacing $\theta(v)$ with a smooth function interpolating between $0$ and $1$, over a time extent characterized by a parameter $v_{t}$, such as $\frac{1}{2}(\tanh v/v_t+1)$. The $v_t\gg1$ regime corresponds to a slow thermalization, which we expect to be well approximated by an adiabatic process (see for example \cite{hubenyrangamanitakayanagi}). We will focus on the thin shell limit ($v_t\rightarrow0$), which describes a fast, non-adiabatic thermalization and allows us to obtain analytic results, returning to comment on the differences between the two descriptions in the final subsection.

Performing the change of coordinates
\begin{equation}\label{eq:tinvr}
t=v-\rho+\frac{\pi}{2},~~\rho\equiv\tan^{-1} r
\end{equation}
we can easily see that $t$ is identified with the static time coordinate $t_i$ inside the shell (i.e. for $v<0$), but not with the static time $t_o$ outside the shell, which makes sense since the Vaidya-BTZ spacetime is not static. In these new coordinates, infalling null geodesics always form angles of $\pi/4$ in $\rho-t$ diagrams, whereas this is true for outgoing null geodesics only in the AdS part $v<0$. Also, $t$ coincides with $v$ on the conformal boundary $S^1\times \mathbb{R}$ at $r\rightarrow\infty$, where it provides a natural time coordinate for the dual field theory and will be denoted by $t_\infty$. Then \eqref{eq:tinvr} implies that the shell starts imploding when $t_\infty=0$.

More details about the structure of this spacetime can be found in \cite{hubenymaxfield}.

\subsection{Geodesic equations}
Following \cite{hubenymaxfield} and exploiting the symmetries of the problem, we can obtain the following first order equations of motion for spacelike geodesics:
\begin{subequations}\label{eq:geodeq}
\begin{align}
L & =r^2\dot{\varphi},\label{eq:geodeq:1}\\
E & =f_\alpha(r)\dot{v}-\dot{r},\label{eq:geodeq:2}\\
\dot{r}^2 & = E^2-\left(\frac{L^2}{r^2}-1\right) f_\alpha(r),\label{eq:geodeq:3}
\end{align}
\end{subequations}
where $\alpha=i,o$ and the dots denote differentiation with respect to an affine parameter $s$ along the geodesics. In the above, $L$ is the angular momentum originating from the spherical symmetry, and $E$ is the energy originating from time translation invariance in $v<0$ and $v>0$. It can be seen that, while $L$ is kept constant along each spacelike geodesic, $E$ is constant in $v<0$ and $v>0$ separately, but is discontinuous at the point where the geodesic crosses the shell, i.e. at $v=0$. In fact, the discontinuity can be computed:
\begin{equation}\label{eq:deltae}
\Delta E\equiv E|_{v=0^+}-E|_{v=0^-}=\frac{1}{2}(f|_{v=0^+}-f|_{v=0^-})\dot{v}|_{v=0}=-\frac{1}{2}(r_+^2+1)\dot{v}|_{v=0},
\end{equation}
with $\dot{v}$ being continuous everywhere.

From the second order equation of motion for $v$
\begin{equation}\label{eq:secondordv}
\ddot{v}=-\frac{1}{2}\frac{\partial f}{\partial r}\dot{v}^2+\frac{L^2}{r^3}
\end{equation}
we observe that $\ddot{v}|_{\dot{v}=0}>0$, which implies that $v$ has one (global) minimum along the geodesic. This will be denoted by $v_{min}$. Now, noting from eq. \eqref{eq:geodeq:2} that flipping the sign of $E$ is equivalent to the reparametrization $s\rightarrow -s$, we will only consider geodesics with $E|_{v_{min}}\geq0$ without loss of generality. Similarly, \eqref{eq:geodeq:1} implies that flipping the sign of $L$ is equivalent to reversing the direction of the increase of $\varphi$, so we will only consider $L\geq0$. Then, eq. \eqref{eq:geodeq:2} and \eqref{eq:geodeq:3} imply that $r|_{v_{min}}=L$.

We are interested in computing the entanglement entropy of boundary theory intervals, the endpoints of which lie at the same boundary time $t_\infty$. In our case, the HRT prescription dictates that this is given by the length of spacelike geodesics with both endpoints anchored on the boundary, at the same time $t_\infty$ (termed ``ETEBA geodesics'' in \cite{hubenymaxfield}). Generally, there are 3 types of ETEBA geodesics in Vaidya-BTZ:
\begin{itemize}
  \item Geodesics lying entirely in the AdS part of the spacetime (i.e. in $v<0$). These must necessarily end at a time slice $t_{\infty}\leq0$ on the boundary.
  \item Geodesics lying entirely in the BTZ part of the spacetime (i.e. in $v>0$). These can only end at $t_{\infty}>0$.
  \item Geodesics that lie in the BTZ part near the boundary, but reach inside the shell. These, too, can only end at $t_{\infty}>0$.
\end{itemize}
Geodesics with $E|_{v_{min}}=0$ have an additional symmetry under $s\rightarrow -s$, so they necessarily end at the same time slice on the boundary (if they do not end at the singularity). In principle we could also have asymmetric ETEBA geodesics, i.e. ETEBA geodesics with $E|_{v_{min}}>0$, but there are good arguments against their existence (see \cite{hubenymaxfield}), so from this point on we will assume $E|_{v_{min}}=0$. Then, geodesics which lie entirely in the AdS or the BTZ part of the spacetime have $E=0$ everywhere, whereas, adopting the convention that $s$ increases in the direction towards the boundary, \eqref{eq:deltae} implies that geodesics which cross the shell have $E=0$ inside the shell and $E<0$ outside.

\subsection{Solutions of the geodesic equations}\label{geosol}
Solving eq. \eqref{eq:geodeq} is straightforward, though somewhat tedious. The details of the calculations, as well as the explicit forms of the geodesics,  can be found in appendix \ref{appendixa}; here we just quote the most important results. In the following, $r_\infty$ is the (large) radial cut-off.
\begin{itemize}
  \item For $t_{\infty}\leq0$ the geometry is pure AdS$_3$ and we obtain the well-known result that, for an interval $\mathcal{I}=[-\varphi_\infty,\varphi_\infty]$ on the boundary, where $\varphi_\infty\in[0,\pi)$, there is a unique spacelike geodesic of length
    \begin{equation}\label{eq:adslength}
    \ell_{AdS}=2\ln(2r_\infty)+2\ln(\sin\varphi_\infty).
    \end{equation}
\end{itemize}
Before moving on to the next cases, we briefly recall what happens in pure BTZ spacetime. As is also explained in \cite{hubenymaxfieldrangamanitonni}, fixing the interval $\mathcal{I}=[-\varphi_\infty,\varphi_\infty]$ on the boundary, the spacelike geodesics can wind around the BTZ black hole an arbitrary number of times. This results from the existence of solutions that have angular extent greater than $2\pi$, but are anchored at the same endpoints $\partial\mathcal{I}$ on the boundary due to the compactness of $\varphi$ (see figure \ref{fig:1fourthbtz} (right)). More precisely, geodesics with angular extent $\varphi^{(k)}\equiv2\varphi_\infty+2\pi k$, $k\in \mathbb{Z}^+$, as well as those with angular extent $\widetilde{\varphi}^{(k)}\equiv-2\varphi_\infty+2\pi(k+1)$, all end on $\partial\mathcal{I}$. However, in the present case the homology constraint is non-trivial; this implies that only $\varphi^{(0)}$ is homologous to $\mathcal{I}$ and $\widetilde{\varphi}^{(0)}$ is homologous to the complementary interval $\bar{\mathcal{I}}$. The lengths of the geodesics are given by $\ell_{BTZ}^{(k)}=2\ln(2r_\infty)+2\ln\left(\frac{1}{r_+}\sinh(r_+\varphi^{(k)})\right)$.

According to the RT prescription, the entanglement entropy of $\mathcal{I}$ is given by the area of the bulk minimal surface anchored on $\partial\mathcal{I}$ and homologous to $\mathcal{I}$. In our case, candidate bulk minimal surfaces include the connected geodesics described above and disconnected surfaces consisting of two pieces: the geodesics and the bifurcation surface of the black hole event horizon $\mathcal{H}$. In the latter case, a geodesic homologous to $\mathcal{I}$, together with $\mathcal{H}$, form an extremal surface which is homologous to $\bar{\mathcal{I}}$. It can be seen that there is a value $\varphi_{\mathcal{\chi}}>\pi/2$ of $\varphi_\infty$ separating two regimes: for $\varphi_\infty<\varphi_{\mathcal{\chi}}$ the HEE is given by the length $\ell_{BTZ}^{(0)}$ of the geodesic with angular extent $\varphi^{(0)}\equiv2\varphi_\infty$, while for $\varphi_\infty>\varphi_{\mathcal{\chi}}$ it is given by the sum of the length $\ell_{BTZ,\bar{\mathcal{I}}}^{(0)}$ of the geodesic with angular extent $\widetilde{\varphi}^{(0)}\equiv2\pi-2\varphi_\infty$ and the bifurcation surface area $\textit{area}(\mathcal{H})$. This regime was termed ``entanglement plateau'' and was explored in \cite{hubenymaxfieldrangamanitonni}.\footnote{See also \cite{b-acs} for related work.} The fact that generally $S_\mathcal{I}\neq S_{\bar{\mathcal{I}}}$ signifies that the BTZ spacetime is dual to a mixed state in the boundary theory (in particular, to a thermal state).

Returning to the study of the Vaidya-BTZ case, we firstly note that, due to the fact that the spacetime is topologically trivial, the homology constraint trivializes (at least for the analogues of the $k=0$ geodesics described above\footnote{More evidence for why $k\neq0$ geodesics don't come into play will be given in subsection \ref{subsection:hee}}). Thus, the disconnected surfaces are absent in our case and we will always have $S_\mathcal{I}=S_{\bar{\mathcal{I}}}$ (signifying that the state is always pure), because $\partial\mathcal{I}\equiv\partial\bar{\mathcal{I}}$ and the geodesics giving the HEEs will be homologous both to $\mathcal{I}$ and $\bar{\mathcal{I}}$. So, from now on the discussion will be focused on a single interval $\mathcal{I}$.
\begin{itemize}
  \item Provided that $t_\infty\geq\varphi_\infty^{(k)}$ (using an obvious notation in analogy with the above), we find geodesics which lie entirely in the BTZ part $v\geq0$\footnote{Note here that the same condition was found in the Poincar\'{e} case (see for example \cite{{allaistonni}}), though only for the solution with $k=0$, since no geodesics wind around the black hole horizon in that case.}. Their lengths are given by:
    \begin{equation}\label{eq:btzlength}
    \ell_{BTZ}^{(k)}=2\ln(2r_\infty)+2\ln\left(\frac{1}{r_+}\sinh(r_+\varphi_\infty^{(k)})\right).
    \end{equation}
    \item Finally, for $t_\infty\geq0$, there exist ETEBA geodesics that cross the shell. Using the radial coordinate at which they encounter the shell $r_s$ (note that they are symmetric, so there is only one such parameter) as a parameter, we find that these geodesics have:
    \begin{subequations}\label{eq:phitotal}
    \begin{equation}\label{eq:phi12}
    \varphi_\infty(r_+,r_s,t_\infty)=\varphi_\infty^{i}(r_+,r_s,t_\infty)+\varphi_\infty^{o}(r_+,r_s,t_\infty),
    \end{equation}
    where $\varphi_\infty^{i}$ is the angular extent inside the shell
    \begin{equation}\label{eq:phiads}
    \varphi_\infty^{i}(r_+,r_s,t_\infty)= \tan^{-1}\left(\frac{2(r_+ - r_s T)} {\sqrt{-4r_+^2-4r_+(r_+^2-1)r_s T+(1+r_+^4+r_+^2(2+4r_s^2))T^2}}\right),
    \end{equation}
    having set $T\equiv\tanh(r_+ t_\infty)$, and
    \begin{equation}\label{eq:phibtz}
    \varphi_\infty^{o}(r_+,r_s,t_\infty)= \frac{1}{r_+}\text{tanh}^{-1}\left(\frac{\sqrt{-4r_+^2-4r_+(r_+^2-1)r_s T+(1+r_+^4+r_+^2(2+4r_s^2))T^2}} {1 - r_+^2 + 2r_+ r_s T}\right)
    \end{equation}
    \end{subequations}
    is the angular extent of the geodesic outside the shell.

    The length is given by the expression
    \begin{subequations}\label{eq:lengthtotal}
    \begin{equation}\label{eq:length12}
    \ell(r_+,r_s,t_\infty)=\ell^{i}(r_+,r_s,t_\infty)+\ell^{o}(r_+,r_s,t_\infty),
    \end{equation}
    with
    \begin{equation}\label{eq:lads}
    \ell^{i}(r_+,r_s,t_\infty)=\ln\left(\frac{(r_+^2+1) T}{(r_+^2+1)T + 4r_s(r_sT-r_+ )}\right),
    \end{equation}
    giving the length of the AdS part and
    \begin{equation}\label{eq:lbtz}
    \ell^{o}(r_+,r_s,t_\infty)=2\ln(2r_\infty)+2\ln\left(\frac{(r_+^2+1)T+2r_s(r_sT-r_+ )}{r_+(r_+^2+1)\sqrt{(1 - T^2)}}\right)
    \end{equation}
    \end{subequations}
    giving the length of the BTZ part.

    Here we should also mention the following constraints, which we obtain by demanding that the geodesics end on the boundary (and not at the singularity):
    \begin{equation}\label{eq:constraints}
    \frac{r_+^2-1+(r_+^2+1)\sqrt{1-T^2}}{2r_+T} \leq r_s \leq \frac{r_+}{T}.
    \end{equation}
    In particular, the first inequality in \eqref{eq:constraints} implies that
    \begin{equation}\label{eq:constraintswot}
    r_s\geq \frac{r_+^2-1}{2r_+},
    \end{equation}
    since $0\leq T\leq1$. The second inequality in \eqref{eq:constraints} trivializes when $r_s\leq r_+$\footnote{These constraints may be thought of as the analogue of the ``critical extremal surface'' of \cite{liusuh1},\cite{liusuh2}. They describe the allowed parameter space for the geodesics to be anchored on the boundary, and also give a significant amount of information about the behavior of certain physical quantities, as will be seen below.}.\\
\end{itemize}

From eq. \eqref{eq:adslength}, \eqref{eq:btzlength} and \eqref{eq:lbtz}, we observe that the divergence in the length of the geodesics appears as the term $2\ln(2r_\infty)$, in which, strictly, we should take the cut-off to infinity $r_\infty\rightarrow\infty$. We will invoke the simplest regularization procedure, which involves subtracting this term from the above equations. From now on, when we refer to the length of the geodesics, we will always mean the regularized length.

\paragraph{Poincar\'{e} limit:}
A good consistency check would be to show that our solutions \eqref{eq:phitotal} and \eqref{eq:lengthtotal} reproduce the known solutions in Poincar\'{e} coordinates, which were first found in \cite{bbbcckmsss}. To go from global to Poincar\'{e} coordinates, effectively we need to ``zoom in'' on the boundary and at the same time make the black hole large (so that it does not disappear). That way, the boundary will look like $\mathbb{R}^{1,1}$ and the horizon will look planar. More precisely, we need to scale
\begin{equation}\label{poincarescaling}
\begin{aligned}
r & \rightarrow\lambda r~(\Rightarrow r_+\rightarrow\lambda r_+,~r_\infty\rightarrow\lambda r_\infty),\\
\varphi & \rightarrow\lambda^{-1}\varphi,\\
v & \rightarrow\lambda^{-1}v~(\Leftrightarrow t\rightarrow\lambda^{-1}t)
\end{aligned}
\end{equation}
and then take the limit $\lambda\rightarrow\infty$. This also has the effect of uncompactifying $\varphi$, since the period becomes infinite. Then, the Vaidya-BTZ metric \eqref{eq:vaidyabtz} takes the form
\begin{equation}\label{eq:vaidyabtzpoincare}
ds^2=-(r^2-\theta(v)r_+^2)dv^2+2dvdr+r^2dx^2,
\end{equation}
in which we have conventionally denoted the uncompactified direction on the boundary by $x$.\\
In this limit, \eqref{eq:phitotal} becomes
\begin{equation}\label{eq:phipoincare}
\varphi_\infty^{P}(r_+,r_s,t_\infty)=\frac{2(r_+-r_sT)}{r_+\sqrt{Tr_+(-4r_s+r_+T+4r_s^2T)}}+
\frac{1}{r_+}\text{tanh}^{-1}\left(\frac{\sqrt{Tr_+(-4r_s+r_+T+4r_s^2T)}}{-r_++2r_sT}\right)
\end{equation}
and \eqref{eq:lengthtotal} becomes
\begin{equation}\label{eq:lengthpoincare}
\ell^P(r_+,r_s,t_\infty)=\ln\left(\frac{4r_\infty^2T(r_+^2T+2r_s(r_sT-r_+ ))^2}{r_+^4(1 - T^2)(r_+^2T + 4r_s(r_sT-r_+ ))}\right).
\end{equation}
We can now see that the solutions obtained in \cite{bbbcckmsss} are equivalent to \eqref{eq:phipoincare} and \eqref{eq:lengthpoincare} after making the appropriate redefinitions.

\paragraph{Graphical representation:}
In principle, we should now invert the expression for the angle $\varphi_\infty$, \eqref{eq:phitotal}, in order to find $r_s(r_+,\varphi_\infty,t_\infty)$ for a fixed $\varphi_\infty$, and then replace it in the expression for the length $\ell(r_+,r_s,t_\infty)$, eq. \eqref{eq:lengthtotal}. This way we can find the evolution of the geodesic length for a fixed interval on the boundary, i.e. $\ell(r_+,\varphi_\infty,t_\infty)$. However, \eqref{eq:phitotal} cannot be solved analytically for $r_s$, so we will solve it numerically in the next section. We end this section by examining some qualitative features of the geodesics in global Vaidya-BTZ.

\begin{figure}
\centering
\includegraphics[width=.80\textwidth]{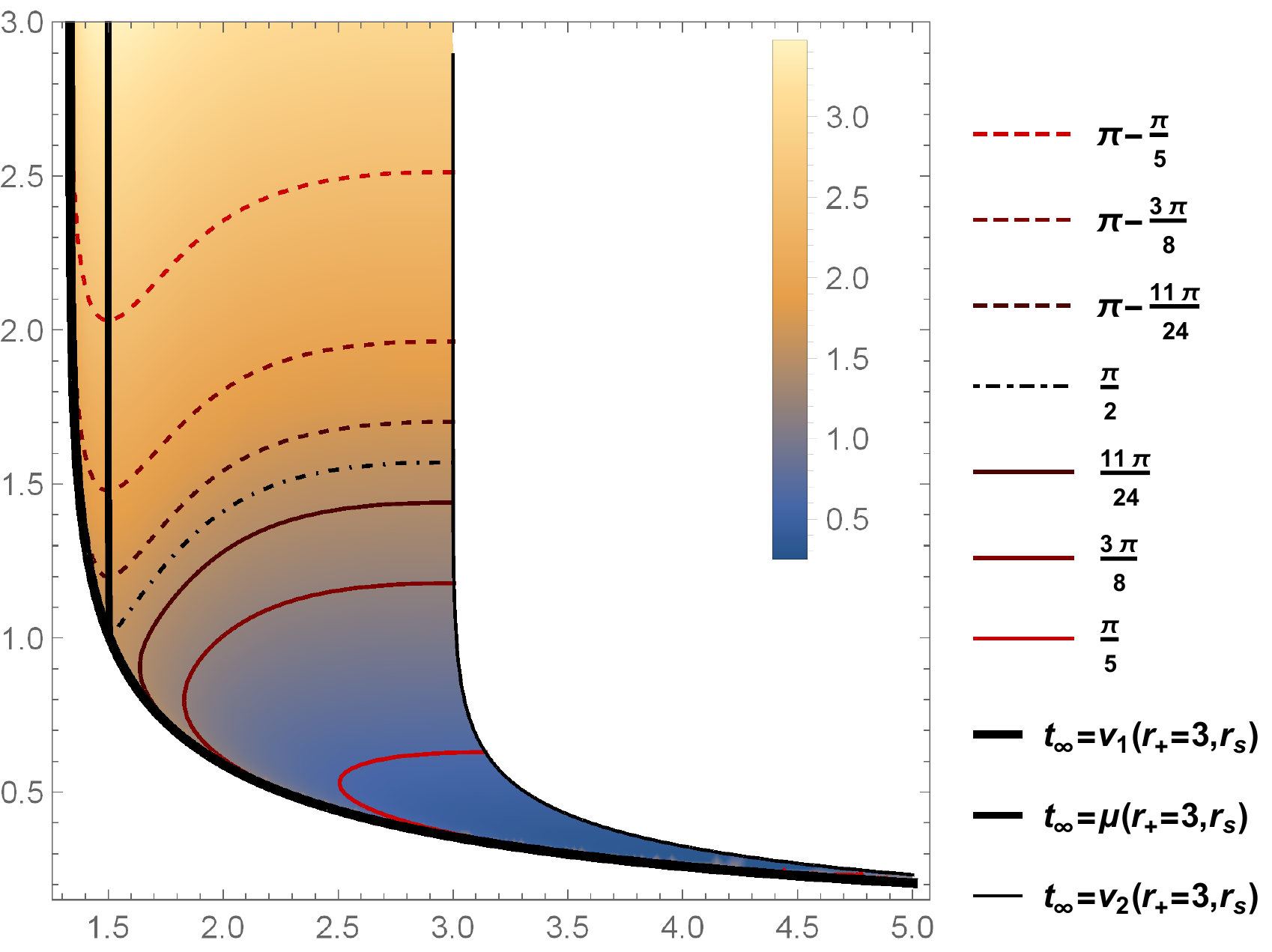}
\begin{picture}(0,0)
\setlength{\unitlength}{1cm}
\put(-13,8){$t_\infty$}
\put(-4.1,-0.2){$r_s$}
\end{picture}
\caption{\label{fig:phi3} Density plot of $\varphi_\infty(r_+=3,r_s,t_\infty)$. The continuous coloured curves are curves of constant $\varphi_\infty$, as can be seen in the legend. The dotted curves ``go around'' the singularity, but they have the same spatial endpoints as the corresponding continuous ones (on time slices specified by $t_\infty$). The dotted-dashed black line, which corresponds to half of the spacetime boundary, can be thought of as a limiting case, with two branches merging into one. The boundaries of the plot are the thick black line $t_\infty=\nu_1(r_+=3,r_s)$ and the thin black line $t_\infty=\nu_2(r_+=3,r_s)$. The curve $\nu_2(3,r_s)$ is a local minimum of $\varphi_\infty$ with $t_\infty$ constant, while $\mu(r_+=3,r_s)$ is a local maximum.}
\end{figure}
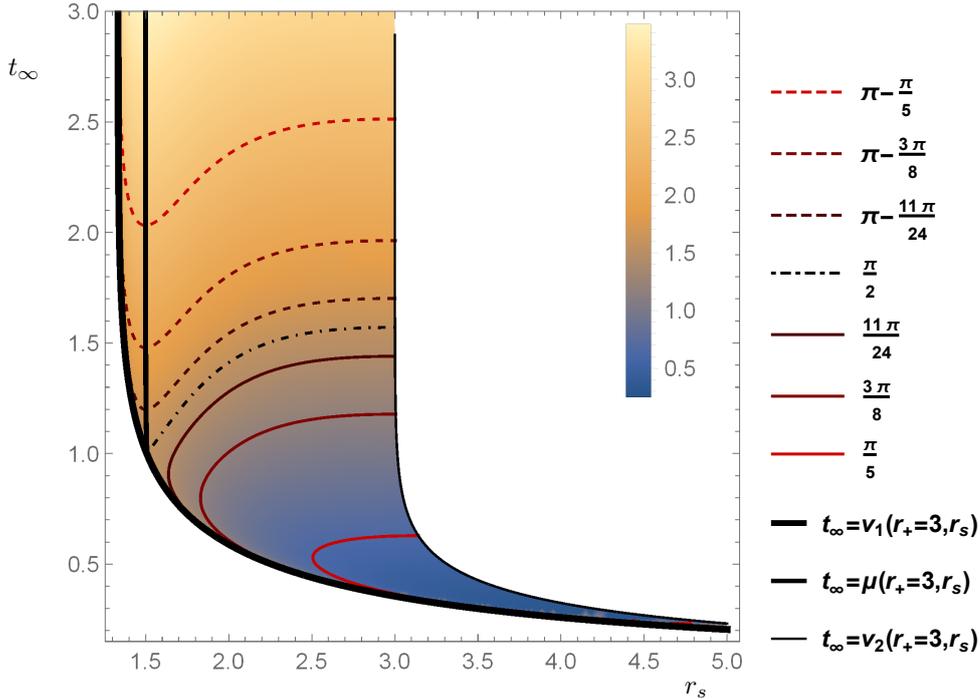
In figure \ref{fig:phi3} we have made the contour plot of $\varphi_\infty(r_+,r_s,t_\infty)$ for a black hole of radius $r_+=3$. The plot continues to $r_s\rightarrow\infty$ and $t_\infty\rightarrow\infty$ in an obvious way. The coloured curves in the $r_s-t_\infty$ plane are curves of constant $\varphi_\infty$. We have also emphasized the boundaries of the plot, which can be read off from constraints \eqref{eq:constraints} for general $r_+$: the thick black line is given by
\begin{equation}\label{eq:nu1}
t_\infty=\nu_1(r_+,r_s)\equiv\frac{1}{r_+}\text{tanh}^{-1}\left(\frac{2r_+((r_+^2-1)r_s + (r_+^2+1)\sqrt{r_s^2+1})}{1 + r_+^4 + 2r_+^2(2r_s^2+1)}\right),
\end{equation}
along which $\varphi_\infty=\frac{\pi}{2}$, and the thin black line is given by
\begin{equation}\label{eq:nu2}
t_\infty=\nu_2(r_+,r_s)\equiv\frac{1}{r_+}\text{tanh}^{-1}\left(\frac{r_+}{r_s}\right),
\end{equation}
marking the time after which the geodesics belonging to a family with a particular angular extent lie entirely in the BTZ part of the spacetime.

\begin{figure}
\centering
\includegraphics[width=.50\textwidth]{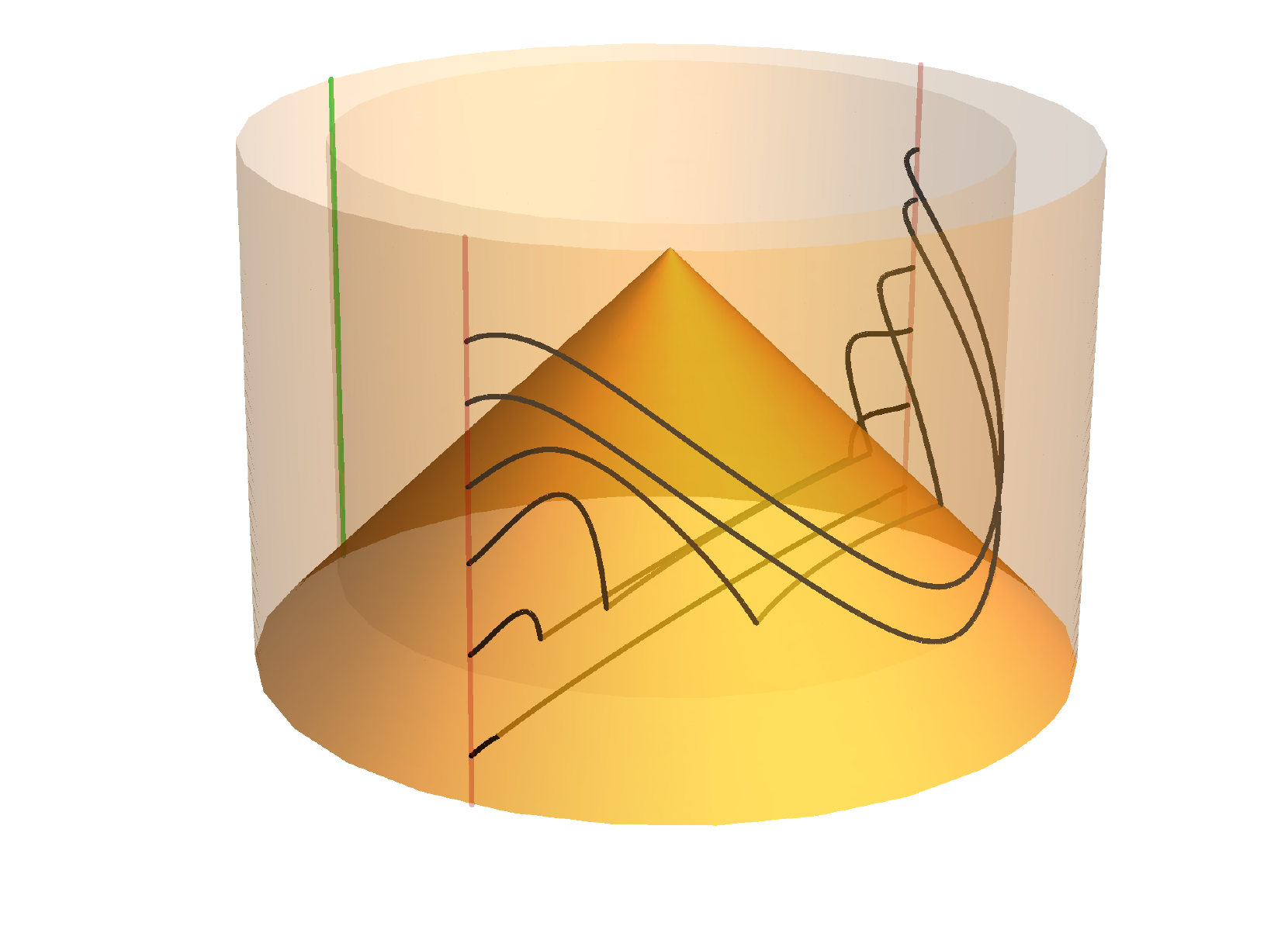}
\hfil
\includegraphics[width=.49\textwidth]{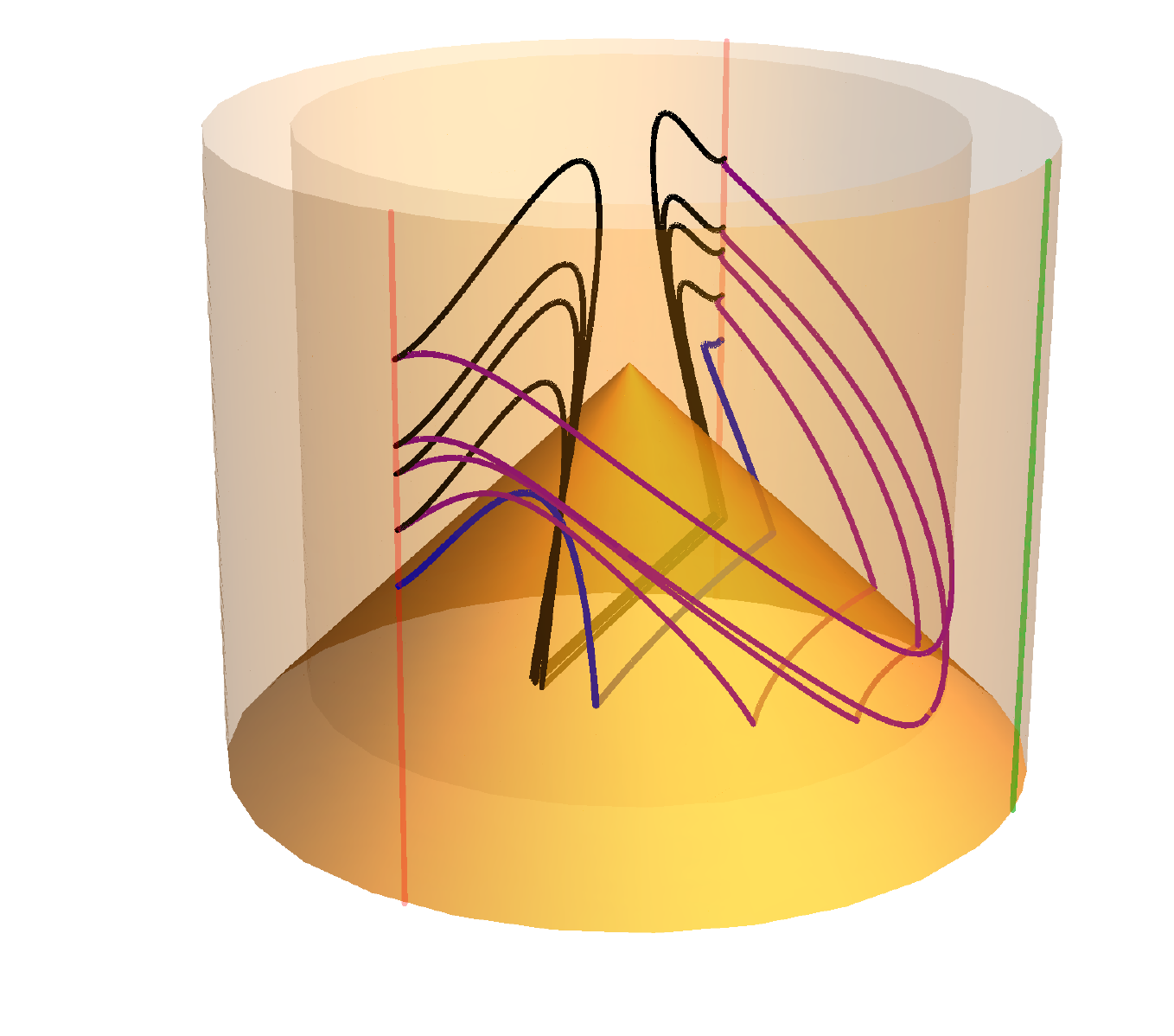}
\caption{\label{fig:1112plot2} In these plots, the time coordinate $t$ runs upwards, and we have compactified the infinite radial direction by plotting $\rho\equiv\tan^{-1}r$. The outer cylinder is the boundary of Vaidya-BTZ, the (null) cone is the incoming shell and the inner cylinder is the event horizon of the black hole formed after the collapse. Here we have chosen $r_+=3$ and we are only plotting the part with $t_\infty\geq0$. Note once again that $t\neq t_o$, and that is why the geodesics lying entirely in the BTZ part of the spacetime do not lie on a constant-t slice. The green lines are reference lines of $\varphi_\infty=\pi$ on the boundary (showing that the plots are rotated with respect to each other). The red lines mark the endpoints of the boundary intervals in question, i.e. $\varphi_\infty=\frac{11\pi}{12}$. \textbf{Left}: Geodesics belonging to the family with angular extent $\frac{11\pi}{12}$. \textbf{Right}: Geodesics belonging to the family with angular extent $2\pi-\frac{11\pi}{12}$. The blue geodesic appears first, which then splits into the ``left branch'' (black) and the ``right branch'' (purple), as shown in the plot.}
\end{figure}

Let us now focus on a fixed interval of length $2\varphi_\infty$ with $\varphi_\infty\leq\frac{\pi}{2}$ on the boundary. We observe that, for $t_\infty\rightarrow0$, we get $r_s\rightarrow\infty$ as expected. Then, as time goes by, $r_s$ gets smaller, reaches a minimum, and grows again, until it reaches the boundary $\nu_2(r_+,r_s)$ of our plot, at which point the geodesic just touches the shell. After that time, the geodesics belonging to that particular family lie entirely in the BTZ part of the spacetime. In figure \ref{fig:1112plot2} (left) we can see this evolution for $\varphi_\infty=\frac{11\pi}{24}$.

It is apparent from figure \ref{fig:phi3} that, at a specific time, a new family of geodesics will come into play. It will have the same endpoints as the family of angular extent $2\varphi_\infty$, but it will have an angular extent of $2\pi-2\varphi_\infty$, i.e. it will ``go around'' the black hole. This new family has two branches of geodesics, separated by the curve
\begin{equation}\label{eq:mu}
t_\infty=\mu(r_+,r_s)\equiv\frac{1}{r_+}\text{tanh}^{-1}\left(\frac{r_+}{2r_s}\right),
\end{equation}
namely the right one, with an evolution and BTZ saturation as the one discussed before, and the left one, which tends asymptotically to the boundary curve $\nu_1(r_+,r_s)$. A graphical representation of the evolution of this family can be seen in figure \ref{fig:1112plot2} (right), again for $\varphi_\infty=\frac{11\pi}{24}$. Note that, for large enough intervals, the second family appears before the first one has saturated to its BTZ form. Finally, it is interesting to observe that the geodesics of the left branch accumulate quickly to the boundary curve $\nu_1(r_+,r_s)$ regardless of the interval size and that $\nu_1$ itself tends to the value $\frac{r_+^2-1}{2r_+}$ for large time $t_\infty$.

For geodesics connecting antipodal points on the boundary, the splitting into radial and non-radial ones is just the degenerate case in which the right branch of the $k=1$ family coincides with a part of the $k=0$ family. This happens at $t|_{cusp}=\frac{1}{r_+}\text{tanh}^{-1}\left(\frac{r_+\sqrt{r_+^2+2}}{r_+^2+1}\right)$, when $\nu_1(r_+,r_s)$ meets $\mu(r_+,r_s)$, at which point the curve $r_s(t_\infty)$ for constant $\varphi_\infty=\pi/2$ is not smooth. We thus understand that there is a spontaneous breaking of the $\mathbb{Z}_2$ reflection symmetry.

In a similar fashion as above, higher-$k$ families, containing geodesics which wind around the black hole multiple times, will come into play as $t_\infty$ increases.

A slight difference that occurs for radii $r_+<1$ is that the curve $\nu_1(r_+,r_s)$ ends on the $t_\infty$ axis at time $\frac{1}{r_+}\text{tanh}^{-1}\left(\frac{2r_+}{r_+^2+1}\right)$ (which is always greater than $t|_{cusp}$). This means that the family of radial geodesics, as well as those belonging to the ``left branch'' of the higher $k$ families, eventually cease to exist (see also the discussion in \cite{hubenymaxfield}).

\section{Holographic entanglement entropy and mutual information}\label{sec:heeandmi}

\subsection{Holographic entanglement entropy}\label{subsection:hee}
The holographic entanglement entropy (HEE) of a boundary interval $\mathcal{I}$ at time $t_\infty$ is given by the length of the minimal ETEBA geodesic anchored on $\partial\mathcal{I}$ at that specific time. In principle, there is an infinite number of geodesics we should take into account, since each family (which is labelled by $k\in\mathbb{Z^+}$, see eq. \eqref{eq:btzlength}), generically contains 2 candidate geodesics, (see figure \ref{fig:1112plot2} (right)), apart from the family with $k=0$ (see figure \ref{fig:1112plot2} (left)). We expect that geodesics which wind around the black hole many times are longer than those which do not, and, although it is hard to prove this directly, eq. \eqref{eq:phitotal}, \eqref{eq:lengthtotal} and \eqref{eq:btzlength} give strong indications that this is the case. We postpone this discussion momentarily, now turning our attention to the $k=0$ families which should give the HEE.

\begin{figure}
\centering
\includegraphics[width=.40\textwidth]{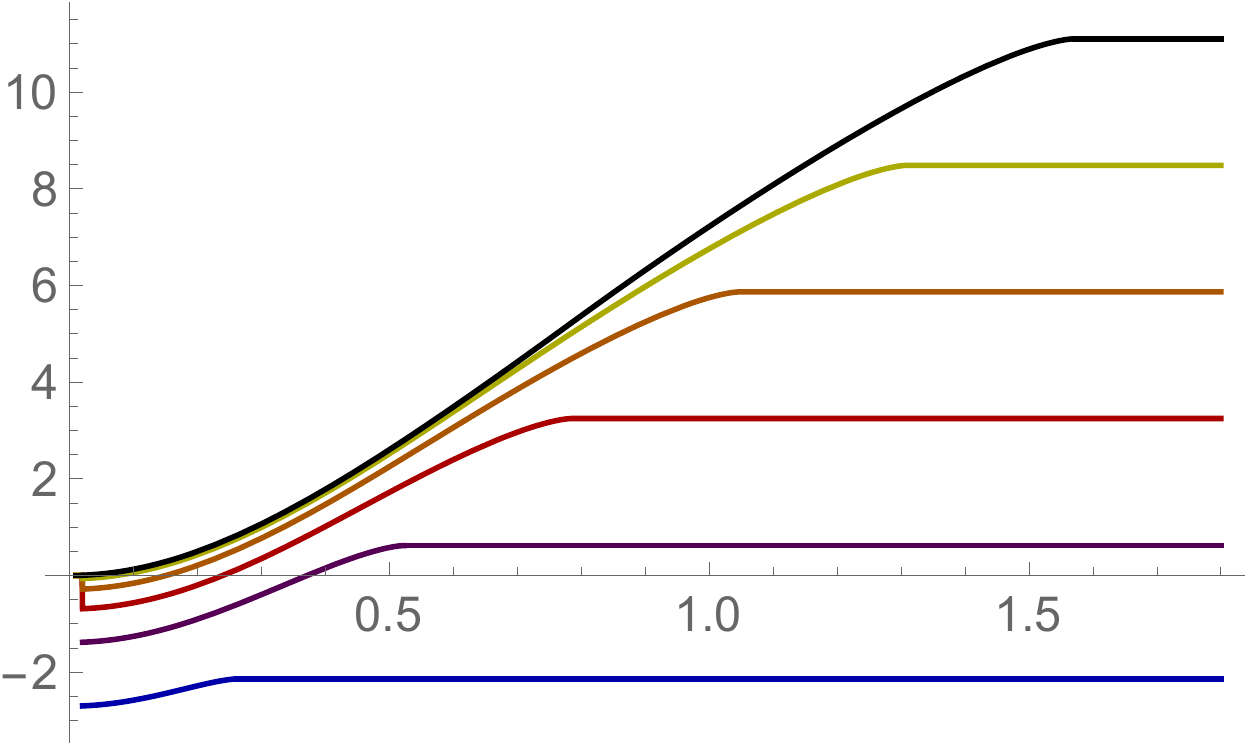}
\hfil
\includegraphics[width=.40\textwidth]{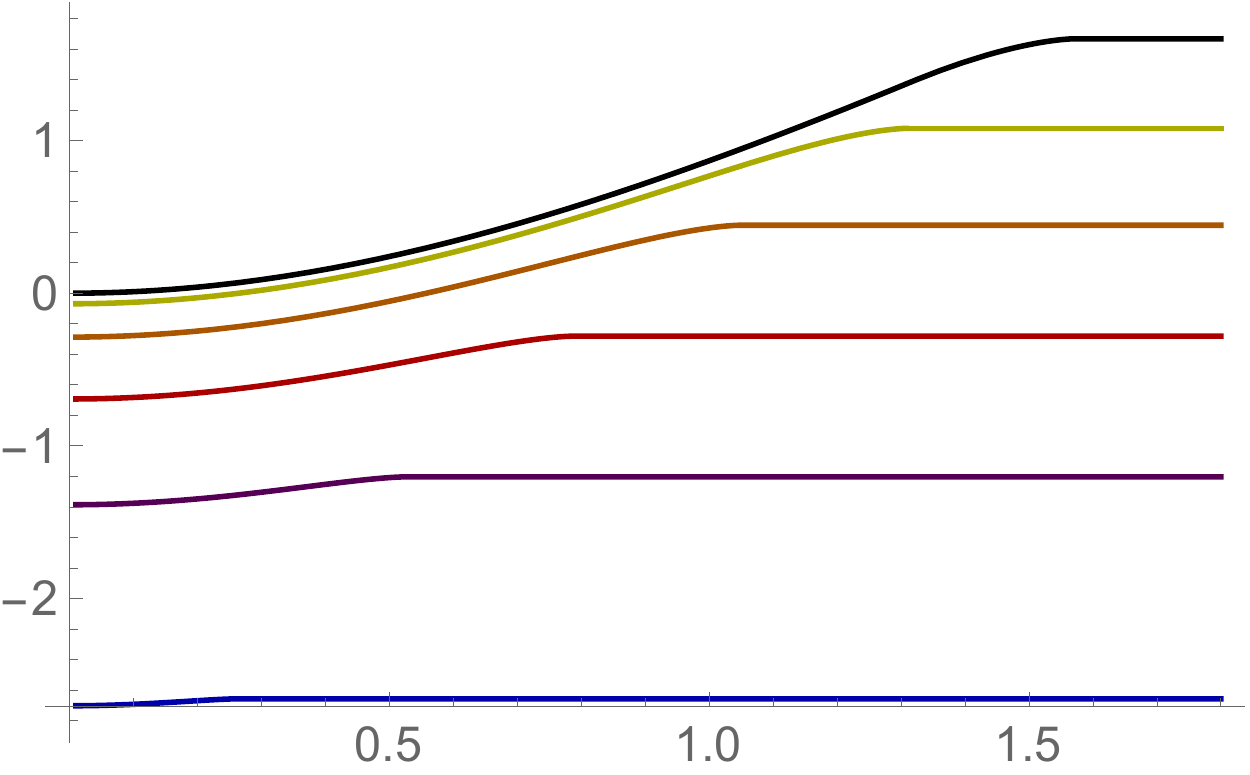}
\hfil
\includegraphics[width=.40\textwidth]{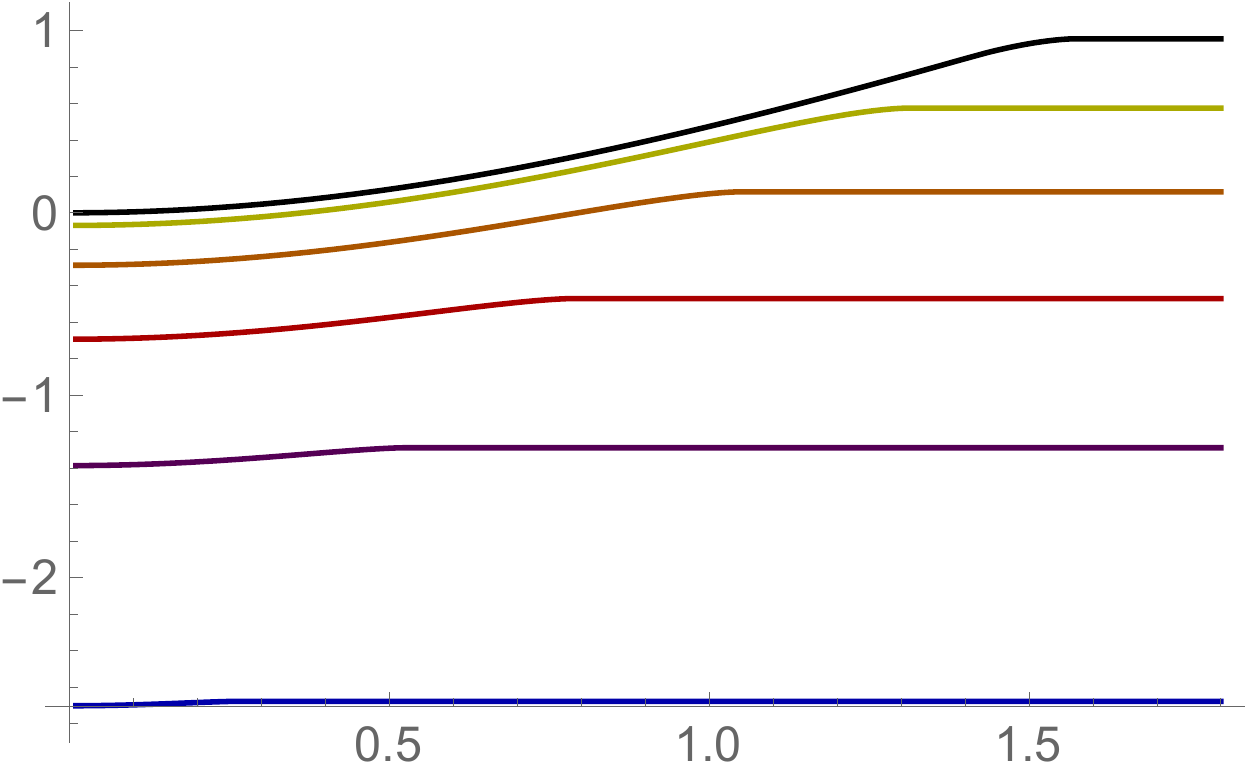}
\begin{picture}(0,0)
\setlength{\unitlength}{1cm}
\put(-7,7.5){$\boxed{r_+=5}$}
\put(0,7.5){$\boxed{r_+=1}$}
\put(-4,3.5){$\boxed{r_+=1/4}$}
\put(-10,7){$\ell$}
\put(-2.8,7){$\ell$}
\put(-6.4,3.2){$\ell$}
\put(3.3,3.7){$t_\infty$}
\put(-0.3,-0.1){$t_\infty$}
\put(-3.6,4.3){$t_\infty$}
\end{picture}
\caption{\label{fig:hee} Holographic entanglement entropy of intervals of total length (from bottom to top) $\pi/6$, $\pi/3$, $\pi/2$, $2\pi/3$, $5\pi/6$, $\pi$, for the cases $r_+=5$ (top left), $r_+=1$ (top right) and $r_+=1/4$ (bottom), as a function of the time on the boundary $t_\infty$.}
\end{figure}
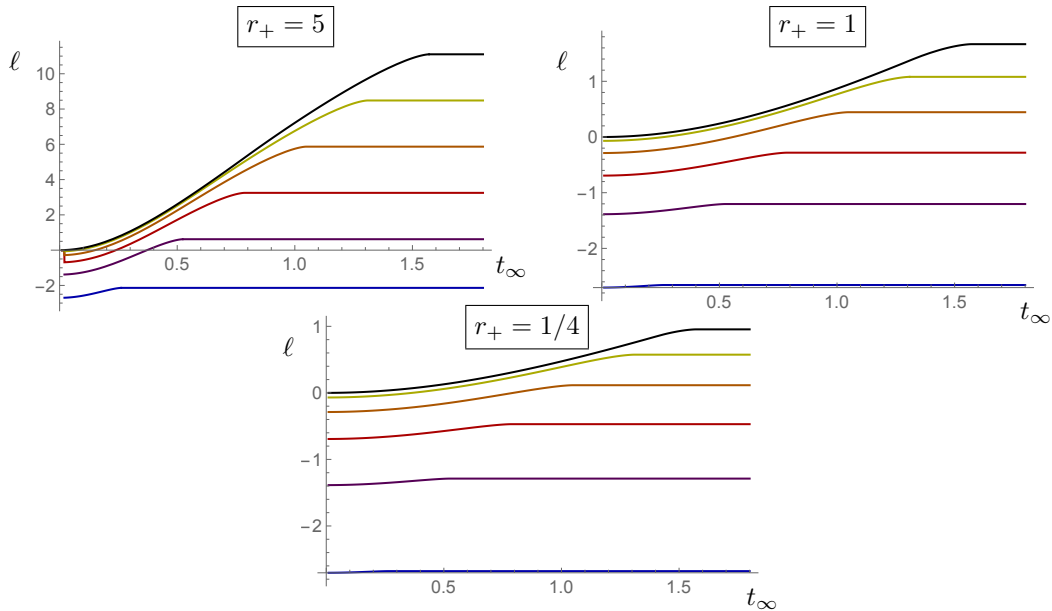

We can numerically invert \eqref{eq:phitotal} to find $r_s(r_+,\varphi_\infty,t_\infty)$ and replace in \eqref{eq:lengthtotal} to express the length as a function $\ell(r_+,\varphi_\infty,t_\infty)$. The resulting plots of the HEE for various interval sizes are presented in figure \ref{fig:hee}, for $r_+=5$ (top left), $r_+=1$ (top right) and $r_+=1/4$ (bottom). We observe that there is a smooth, monotonic interpolation between the AdS value and the BTZ value, always increasing since eq. \eqref{eq:adslength} and \eqref{eq:btzlength} imply that $\ell_i<\ell_o$ for all $\varphi_\infty$. As expected from \eqref{eq:adslength}, the initial value of the HEE for intervals of length $\pi/2$ is 0, while it goes to $-\infty$ as the length goes to 0\footnote{Note that we are only considering the constant part of the HEE, having ignored the term $2\log(2r_\infty)$ which diverges quicker.}. Importantly, the linear growth regime initially predicted from general CFT arguments in \cite{calabresecardy1}, and verified from a holographic viewpoint in \cite{hartmanmaldacena}--\cite{bbbcckmsss2}, \cite{lwwy}--\cite{liusuh2} and \cite{knstv} (among others), is absent in our case. This is not surprising: the derivation of \cite{calabresecardy1} relied crucially on defining the 1+1 dimensional quantum field theory on $\mathbb{R}^{1,1}$, as well as starting from a gapped theory before the quantum quench, and ending with a CFT after the quench (but it was independent of the coupling). Physically, this means that initially there were only short range correlations at a scale set by the inverse mass gap, inside which the proposed quasiparticles that ``carry the entanglement'' are produced, and then travel freely to infinity with a velocity $v=1$. This toy model explains the linear growth and saturation found in that case. In the context of AdS/CFT, initially (i.e. for $t_\infty<0$), the boundary theory state is the vacuum of a 1+1 dimensional CFT, since it is dual to pure AdS$_3$, and so we start with long range correlations. Despite this difference, the linear growth regime can also be derived from the gravity perspective by using the separation of scales $r_+\varphi_\infty\gg r_+t_\infty\gg 1$ (see for example \cite{liusuh2}). The effects of this separation of scales will start appearing as $r_+$ becomes larger, and indeed it can be observed in figure \ref{fig:hee} that a linear regime starts forming when $r_+=5$.

Although the free quasiparticle model is adequate to capture the growth of the entanglement entropy of one interval after a global quench, it fails when the subsystem consists of two or more disconnected components, as was noted in \cite{leichenauermoosa}: it predicts a non monotonic evolution, while the holographic prescription and large $c$ CFT calculations show that it increases until it reaches its equilibrium value. The same contradiction arises for our compact system: the periodic, sawtooth evolution predicted by the above toy model is in sharp contrast to the plots in figure \ref{fig:hee}. We thus expect that figure \ref{fig:hee} captures the essential characteristics of global quenches in compact 1+1 dimensional quantum systems, at least for large central charge $c$\footnote{Indeed, it was recently shown in \cite{abgh} that a small enough central charge $c$ leads to an HEE profile in agreement with the quasiparticle model of \cite{calabresecardy1}, while in the holographic regime ($c\gg1$, strong coupling), the profile agrees with the one predicted using the ``HRT prescription''.}.

A simple calculation of the behavior of the HEE at early and late times (see appendix \ref{appendixb}) confirms the findings of \cite{hubenyrangamanitonnitonni}: for $t_\infty\rightarrow0$ we find
\begin{equation}\label{eq:earlytimehee}
\ell(\varphi_\infty,r_+,t_\infty)=2\ln(\sin(\varphi_\infty))+\frac{r_+^2+1}{2}t_\infty^2+\mathcal{O}(t_\infty^3),
\end{equation}
while for $t_\infty\rightarrow\varphi_\infty$
\begin{equation}\label{eq:latetimehee}
\begin{aligned}
\ell(\varphi_\infty,r_+,t_\infty) & =2\ln\left(\frac{1}{r_+}\sinh(r_+\varphi_\infty)\right)-\frac{2\sqrt{2}(r_+^2+1)\sqrt{\tanh(r_+\varphi_\infty)}}{3\sqrt{r_+}} (\varphi_\infty-t_\infty)^{3/2}\\
& +\mathcal{O}((\varphi_\infty-t_\infty)^2).\\
\end{aligned}
\end{equation}

\paragraph{$k>0$ families:}
Returning to figure \ref{fig:phi3}, we make the following observations which hold for general $r_+$:
Firstly, eq. \eqref{eq:btzlength} shows that $\ell_o^k$ is an increasing function of $k$. Secondly, it can be checked that the qualitative features of $\varphi_\infty(r_+,r_s,t_\infty)$ agree with those of $\ell(r_+,r_s,t_\infty)$; in particular, the curve $\nu_2$ is a (local) minimum and $\mu$ is a maximum (for constant $r_s$), while the parameter space is of course bounded by $\nu_1$ and $\nu_2$. It can be shown by a direct calculation that the length $\ell$ increases along curves $r_s(t_\infty)$ of constant $\varphi_\infty$, as well as along $\nu_1$, $\nu_2$ and $\mu$ (see appendix \ref{appendixb}).
In fact, along $\nu_2$ the length is
\begin{equation}\label{eq:lalongnu2}
\ell(r_+,t_\infty)=2\ln\left(\frac{1}{r_+}\sinh(r_+t_\infty)\right)
\end{equation}
and along $\nu_1$ the length is
\begin{equation}\label{eq:lalongnu1}
\ell(r_+,t_\infty)=2\ln\left(\frac{-1+r_+^2+(1+r_+^2)\cosh(r_+t_\infty)}{2r_+^2}\right)\sim2r_+t_\infty,
\end{equation}
for $r_+t_\infty\gg1$. It is interesting to note here that in the $r_+<1$ case, after $\nu_1(r_+,r_s)$ terminates, the length along the $r_s=0$ axis is also given by eq. \eqref{eq:lalongnu2} above.

In principle, it is possible that for some $t_\infty$, the geodesics belonging to the $k=0$ and $k=1$ families exchange dominance, i.e. the minimal geodesics belong to the $k=1$ family. However, the allowed parameter space for this to happen can be significantly restricted by the above observations. Firstly, note that we can show that the length increases along the $k=0$ families of fixed $\varphi_\infty$, as well as along the left and the right branch of $k\geq1$ families (see appendix \ref{appendixb}). Also, since for constant $t_\infty$ the length is monotonically decreasing from $\mu$ to $\nu_2$, we can see that when they first appear, the higher $k$ families will never be shorter than those belonging to lower $k$ families, and this will remain so for the right branch geodesics. This confirms the intuition that geodesics winding around the black hole are always longer than those which don't. Regarding the left branch, which contains geodesics reaching deep inside the shell, we observe from figure \ref{fig:phi3} that they accumulate quickly near the boundary curve $\nu_1$, which contains geodesics with lengths growing almost linearly with time $t_\infty$ (see eq. \eqref{eq:lalongnu1}). So we expect them to be even longer, although we cannot rule out the possibility that something counterintuitive happens at the beginning of the left branch. Similar arguments can be applied to families with general $k\geq1$\footnote{Note also that in some cases the geodesics may even be excluded from the possibility of giving the HEE due to the homology constraint.}.

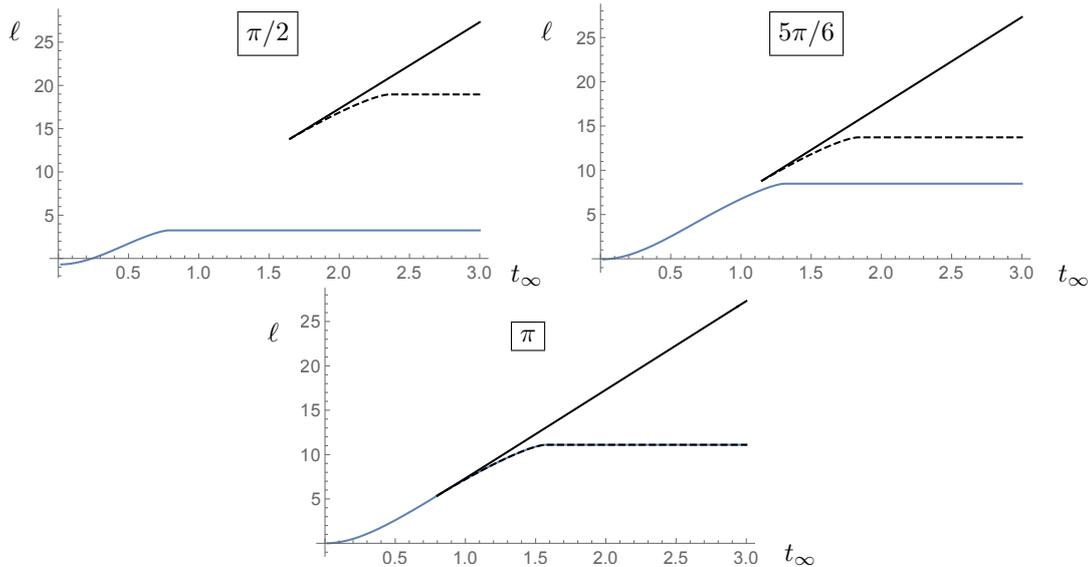
\begin{figure}
\centering
\includegraphics[width=.40\textwidth]{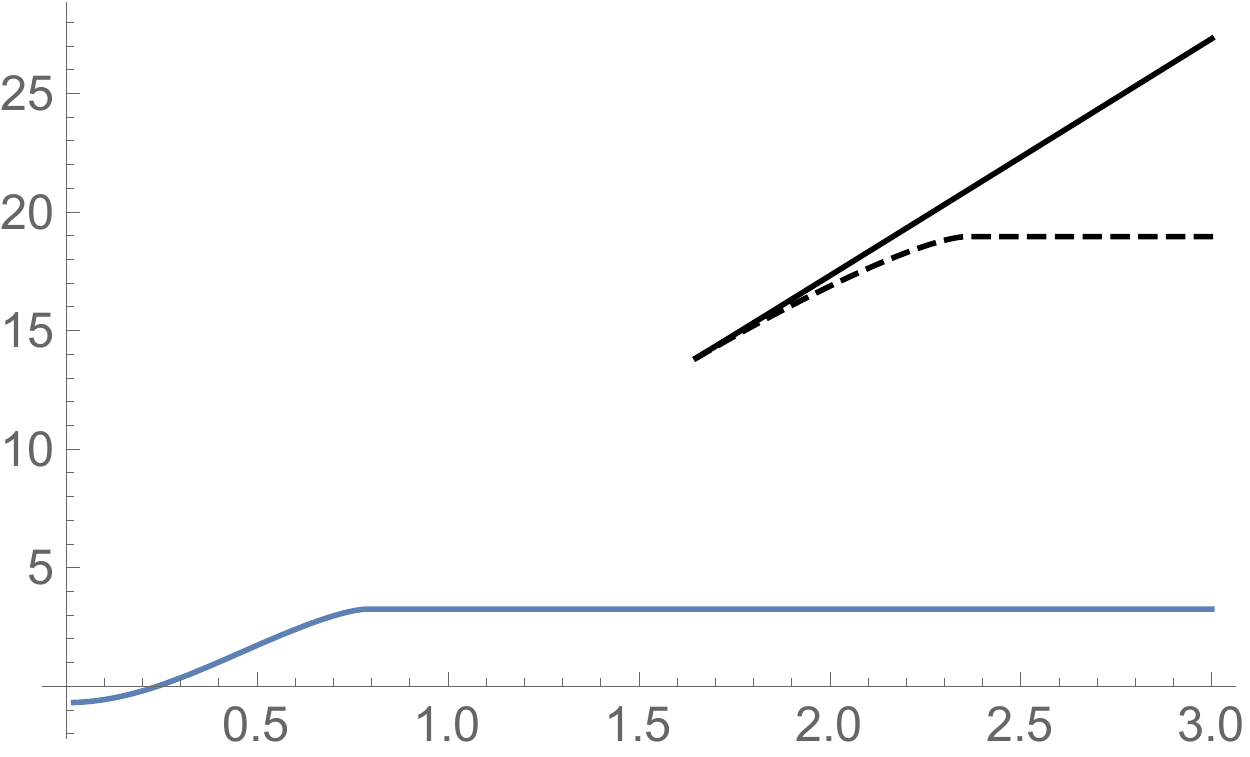}
\hfil
\includegraphics[width=.40\textwidth]{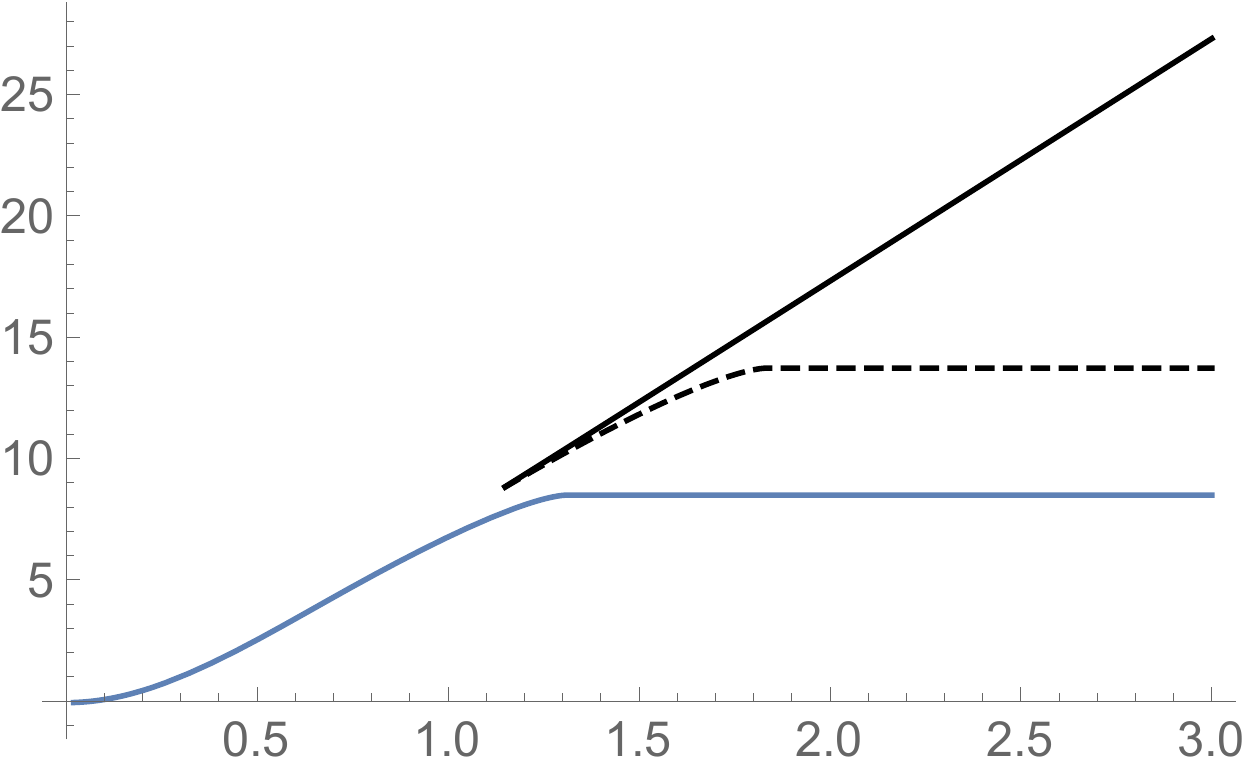}
\hfil
\includegraphics[width=.40\textwidth]{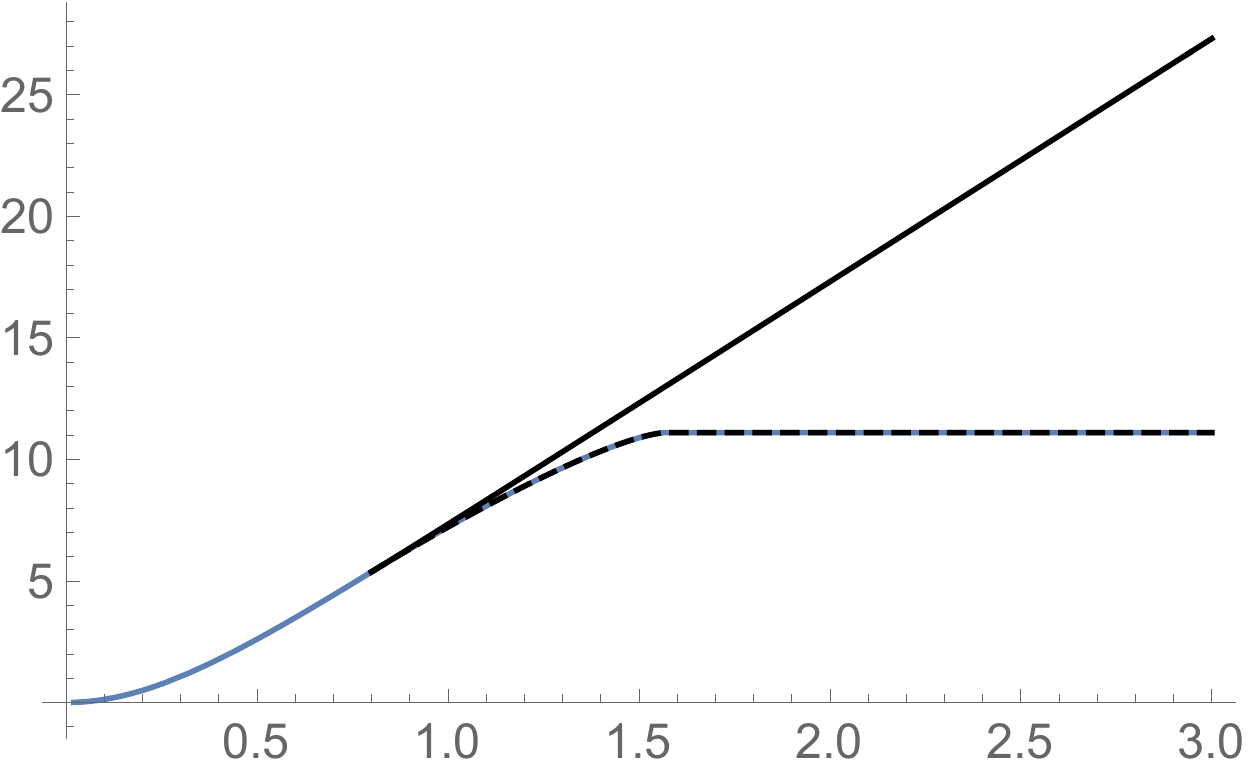}
\begin{picture}(0,0)
\setlength{\unitlength}{1cm}
\put(-10,7){$\ell$}
\put(-3,7){$\ell$}
\put(-6.6,3){$\ell$}
\put(3.8,3.8){$t_\infty$}
\put(0.2,0.1){$t_\infty$}
\put(-3.4,3.8){$t_\infty$}
\put(-7,7){$\boxed{\pi/2}$}
\put(0,7){$\boxed{5\pi/6}$}
\put(-3.4,3){$\boxed{\pi}$}
\end{picture}
\caption{\label{fig:fam} Time evolution of holographic entanglement entropy of $k=0$ (blue), and the left branch (black) and right branch (black, dashed) of $k=1$ families of geodesics. The radius of the black hole is $r_+=5$ and the intervals on the boundary are of total length: $\pi/2$ (top left), $5\pi/6$ (top right), and $\pi$ (bottom).}
\end{figure}

After an extensive numerical search of the parameter space, the general behavior is found to agree with our intuition, and can be seen in figure \ref{fig:fam}. Thus, the saturation to equilibrium is found to be continuous.

\paragraph{Rate of growth:}

An interesting quantity to compute is the dimensionless rate of growth
\begin{equation}\label{rateofgrowth}
\mathfrak{R}(t_\infty)\equiv\frac{1}{s_{eq} A}\frac{\mathrm{d} S_\mathcal{I}}{\mathrm{d} t_\infty},
\end{equation}
introduced in \cite{liusuh1} and \cite{liusuh2} for a boundary region $\mathcal{I}$ in a general $d$-dimensional boundary. In the above, $A$ is the entangling surface of $\mathcal{I}$, i.e. $A=\textit{area}(\partial\mathcal{I})$ and $s_{eq}$ is the equilibrium entropy density. $s_{eq}$ is defined such that the difference of the equilibrium entropy from the vacuum entropy $\Delta S_{eq}$ takes the form $\Delta S_{eq}\approx s_{eq} \textit{vol}(\mathcal{I})$ for regions $\mathcal{I}$ with large effective size $R_{eff}$ compared to the equilibrium scale $1/r_+$ set by the radius of the black hole. The definition of the rate of growth was motivated from the fact that, in a very large class of Vaidya-type geometries with \textit{non-compact} boundary, there exists a universal ``post-local-equilibration linear growth'' of the form
\begin{equation}\label{plelineargrowth}
S_\mathcal{I}(t_\infty)=v_E s_{eq} A t_\infty + \ldots,
\end{equation}
where $v_E$ is a shape-independent dimensionless quantity. This linear growth regime appears when $R_{eff}\gg t_\infty \gg 1/r_+$. When $t_\infty \ll 1/r_+$ a quadratic ``pre-local-equilibration growth'' was observed.


Specializing to the Vaidya-BTZ case in Poincar\'{e} coordinates and translating to our notation, we get: $s_{eq}=r_+$\footnote{In our analysis we consistently ignore the factor of $1/4G_N$, as explained in the introduction.}, $A=2$ (reflecting the fact that a boundary interval has 2 endpoints) and $v_E=1$ (a universal result for the linear growth regime in non-compact 1+1 dimensional CFTs). So we get
\begin{subequations}\label{eq:prepost2d}
\begin{align}
\Delta\ell & =\frac{r_+^2}{2}t_\infty^2,~~r_+t_\infty\ll1\label{eq:prepost2d:1}\\
\Delta\ell & =2r_+t_\infty,~~r_+\varphi_\infty\gg r_+t_\infty\gg1\label{eq:prepost2d:2}
\end{align}
\end{subequations}
and
\begin{equation}\label{eq:ratemycase}
\mathfrak{R}^{(2)}(t_\infty)=\frac{1}{2r_+}\frac{\mathrm{d} \ell}{\mathrm{d} t_\infty}.
\end{equation}
We also note that extensivity for the equilibrium entropy is derived by the expression for the length of a spacelike geodesic in BTZ spacetime: eq. \eqref{eq:btzlength} implies $\ell_{BTZ}\approx r_+(2\varphi_\infty)$ when $r_+\varphi_\infty\gg1$.

Natural arguments led the authors to suggest that $\mathfrak{R}\leq1$ for the class of systems they studied and in appendix \ref{appendixc} it is shown that this holds in the Vaidya-BTZ case in Poincar\'{e} coordinates. However, it is easy to see that we can have $\mathfrak{R}^{(2)}>1$ in the case of global coordinates for small enough $r_+$\footnote{Numerical calculation of $\mathfrak{R}^{(2)}$ for various black hole radii $r_+$ and boundary interval sizes $\varphi_\infty$ is in agreement with the following arguments.}. For example, we can compute the rate of growth of a boundary interval with $\varphi_\infty=\pi/2$ at the time when the geodesics stop being radial: the time derivative of eq. \eqref{eq:lalongnu1} evaluated at time $t|_{cusp}=\frac{1}{r_+}\text{tanh}^{-1}\left(\frac{r_+\sqrt{r_+^2+2}}{r_+^2+1}\right)$ gives
\begin{equation}\label{eq:ratepi2cusp}
\mathfrak{R}^{(2)}|_{cusp}=\frac{1}{2r_+}\frac{2(r_+^2+1)\sqrt{r_+^2+2}}{r_+^2+3}.
\end{equation}
This implies that $\mathfrak{R}^{(2)}|_{cusp}\geq1$ for $r_+\leq\sqrt{\sqrt{2}-1}<1$, while $\mathfrak{R}^{(2)}|_{cusp}\rightarrow1$ from below quickly as $r_+$ grows larger.

This does not imply any violation of causality, but only that the definition \eqref{eq:ratemycase} ceases to have a nice interpretation as the rate of growth of HEE when $r_+$ is small enough. More precisely, a small $r_+$ cannot be interpreted as the entropy density $s_{eq}$ if our system is compact, since extensivity will fail even for the biggest interval that we may consider (we necessarily have $\varphi_\infty\lesssim\mathcal{O}(1)$). Indeed, we see that the divergence at $r_+=0$ comes from the normalization $1/2r_+$. This also makes sense if we recall that the extensive behavior of the equilibrium HEE comes from the part of the geodesic which ``lies along the horizon'', so a small radius black hole cannot have this effect in a compact boundary spacetime. On the other hand, in the flat boundary case, $r_+$ can always be interpreted as the entropy density $s_{eq}$, since we can consider arbitrarily large boundary intervals, able to compensate for the small $r_+$. Geometrically, a spacelike geodesic anchored on a large enough boundary interval will always contain a part lying along a flat, non-compact black hole horizon.

\subsection{Holographic mutual information}\label{subsection:hmi}

We now turn to the evolution of the holographic mutual information, where we can observe important deviations from the adiabatic approximation and a generalization of the previous findings in the Poincar\'{e} case (\cite{bbccg} and \cite{allaistonni}). All qualitatively different behaviors are presented in figure \ref{fig:hmi} below, for $r_+=5$ (without loss of generality).

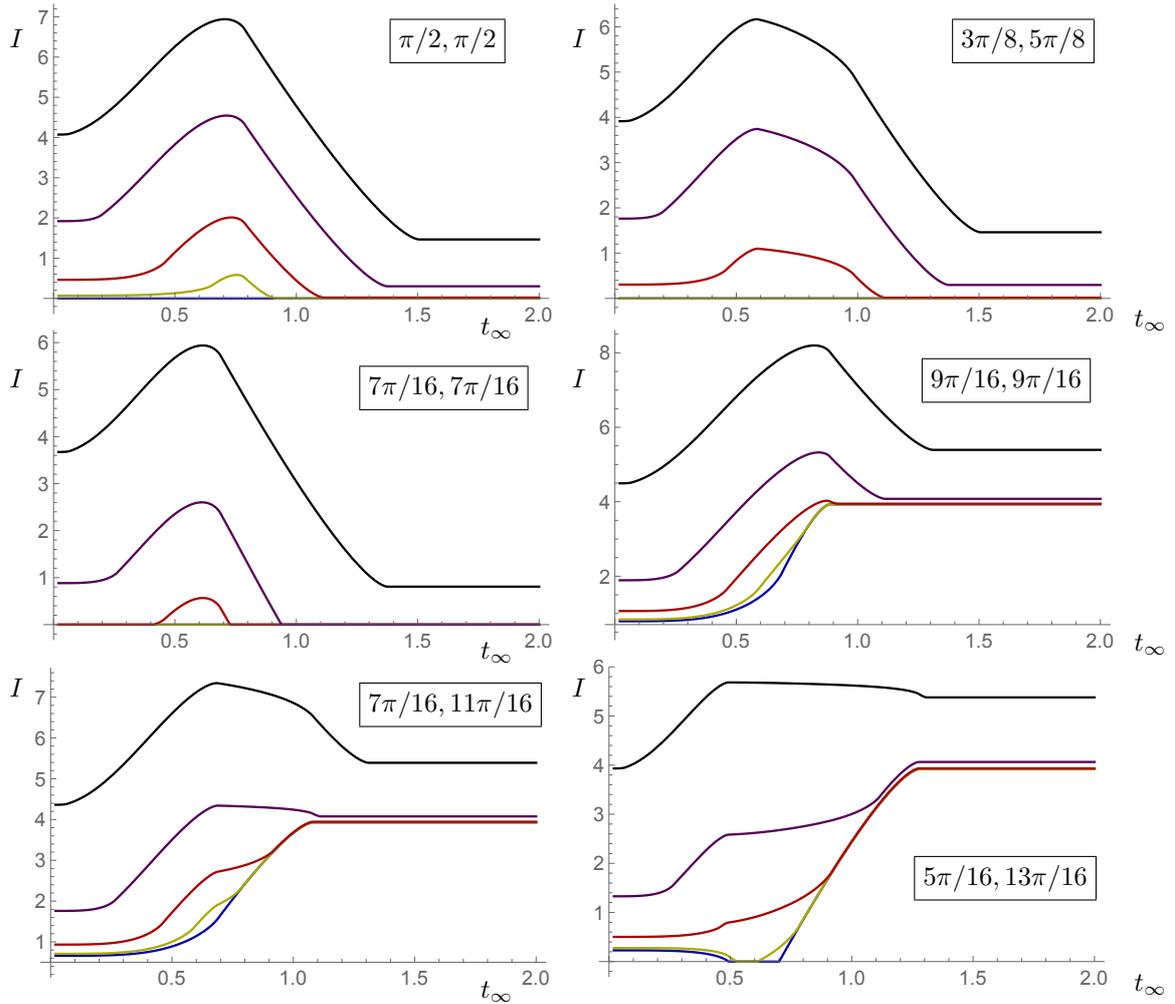
\begin{figure}
\centering
\includegraphics[width=.45\textwidth]{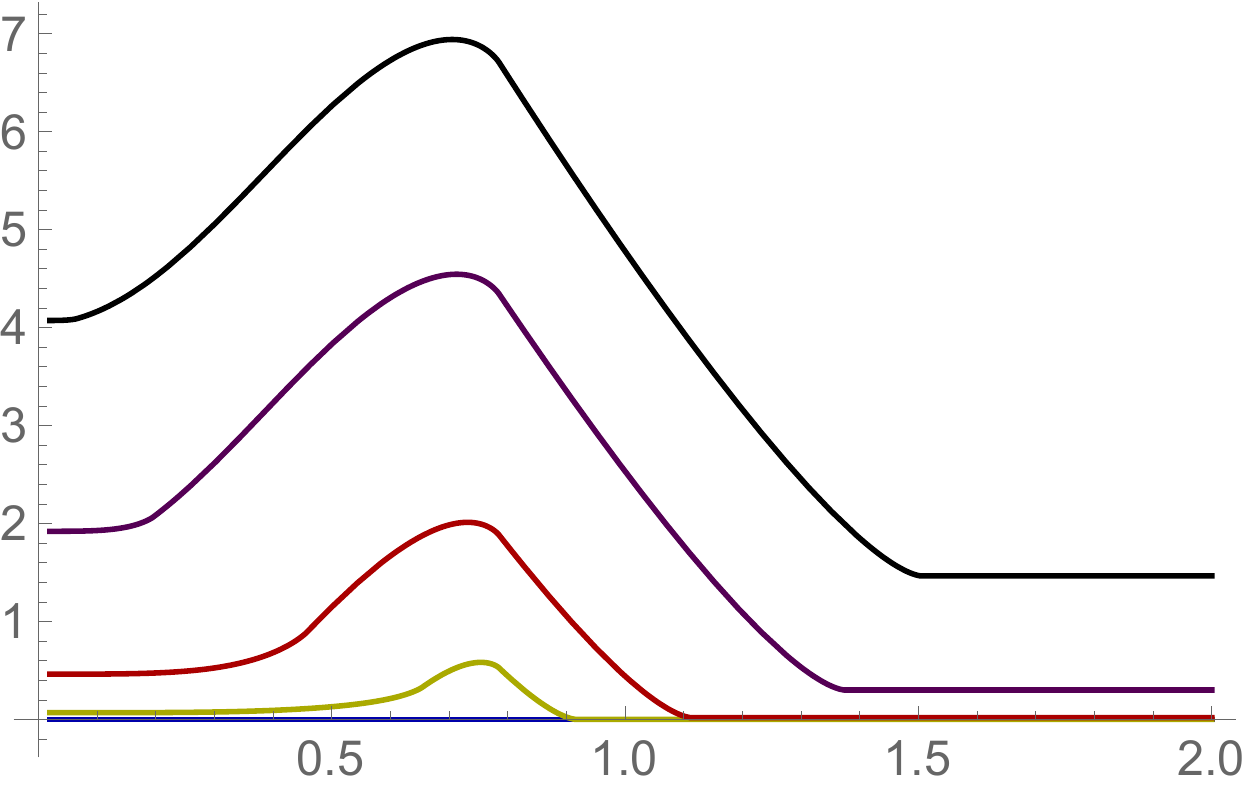}
\hfil
\includegraphics[width=.45\textwidth]{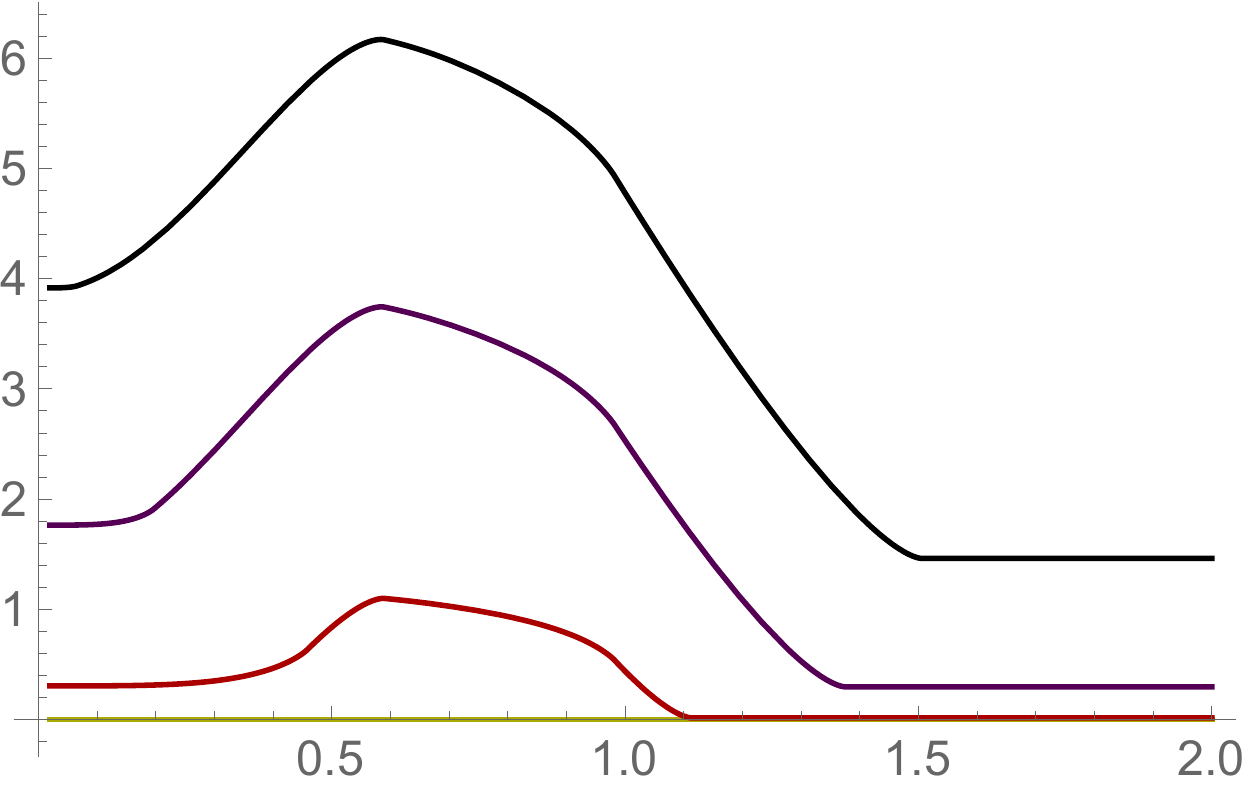}
\hfil
\includegraphics[width=.45\textwidth]{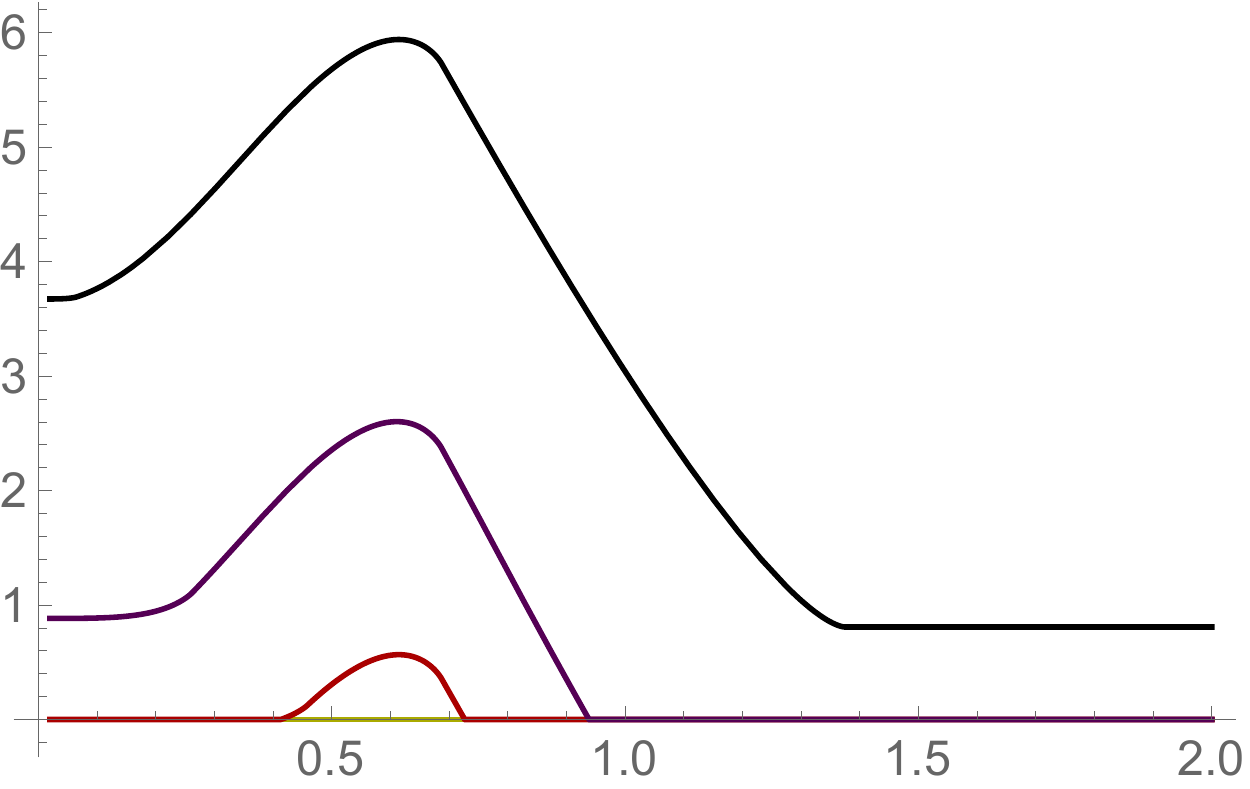}
\hfil
\includegraphics[width=.45\textwidth]{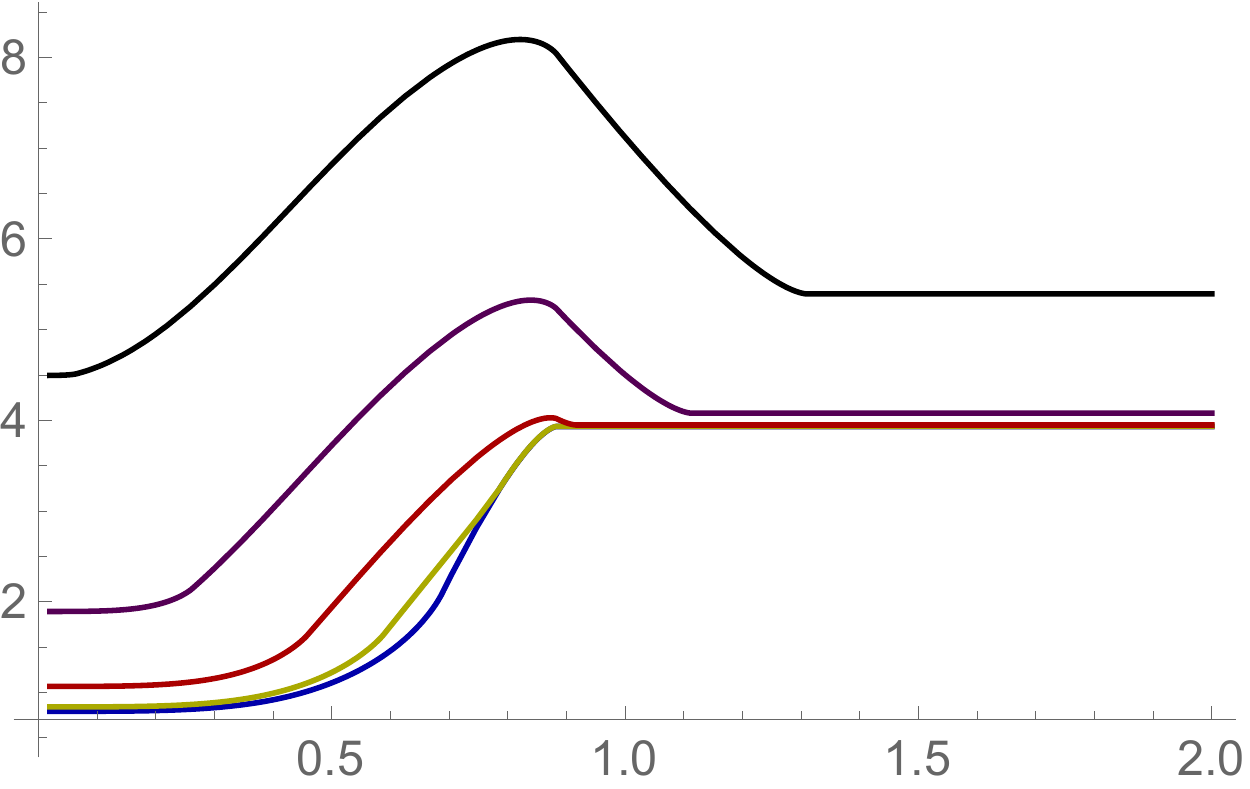}
\hfil
\includegraphics[width=.45\textwidth]{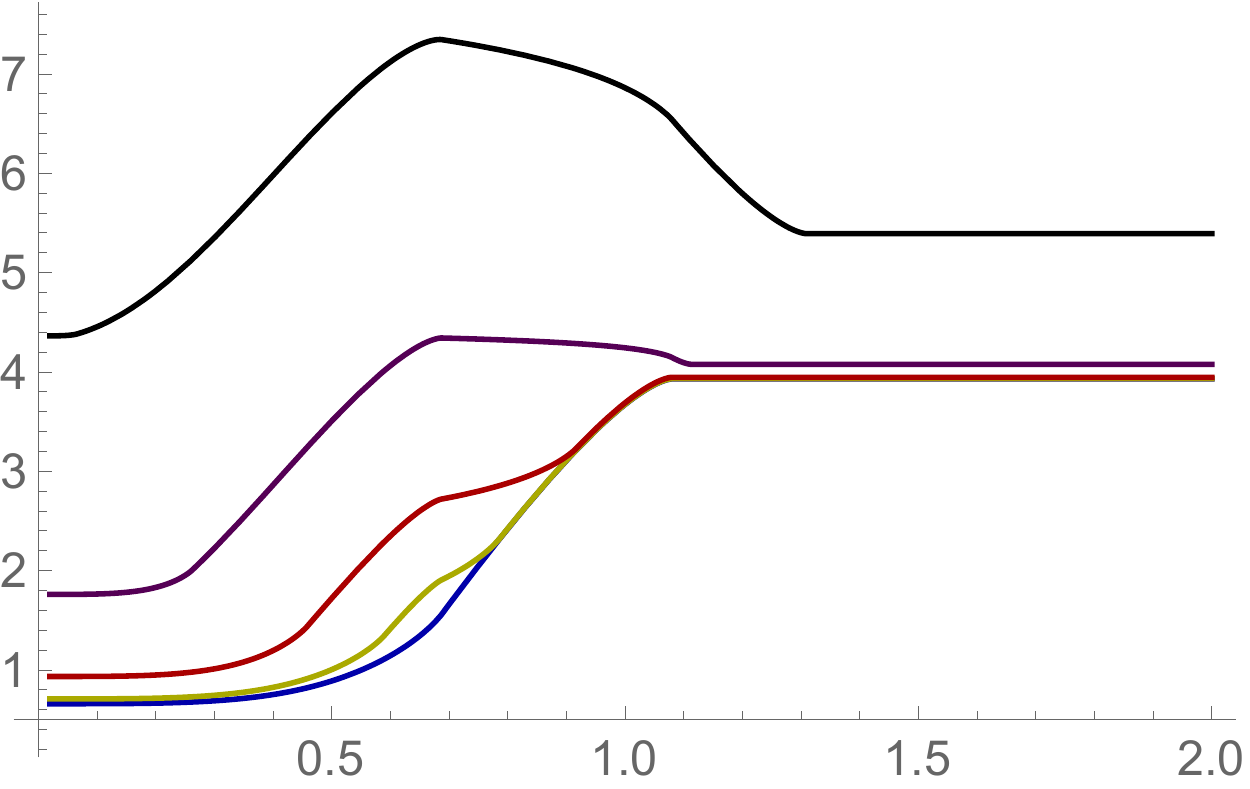}
\hfil
\includegraphics[width=.45\textwidth]{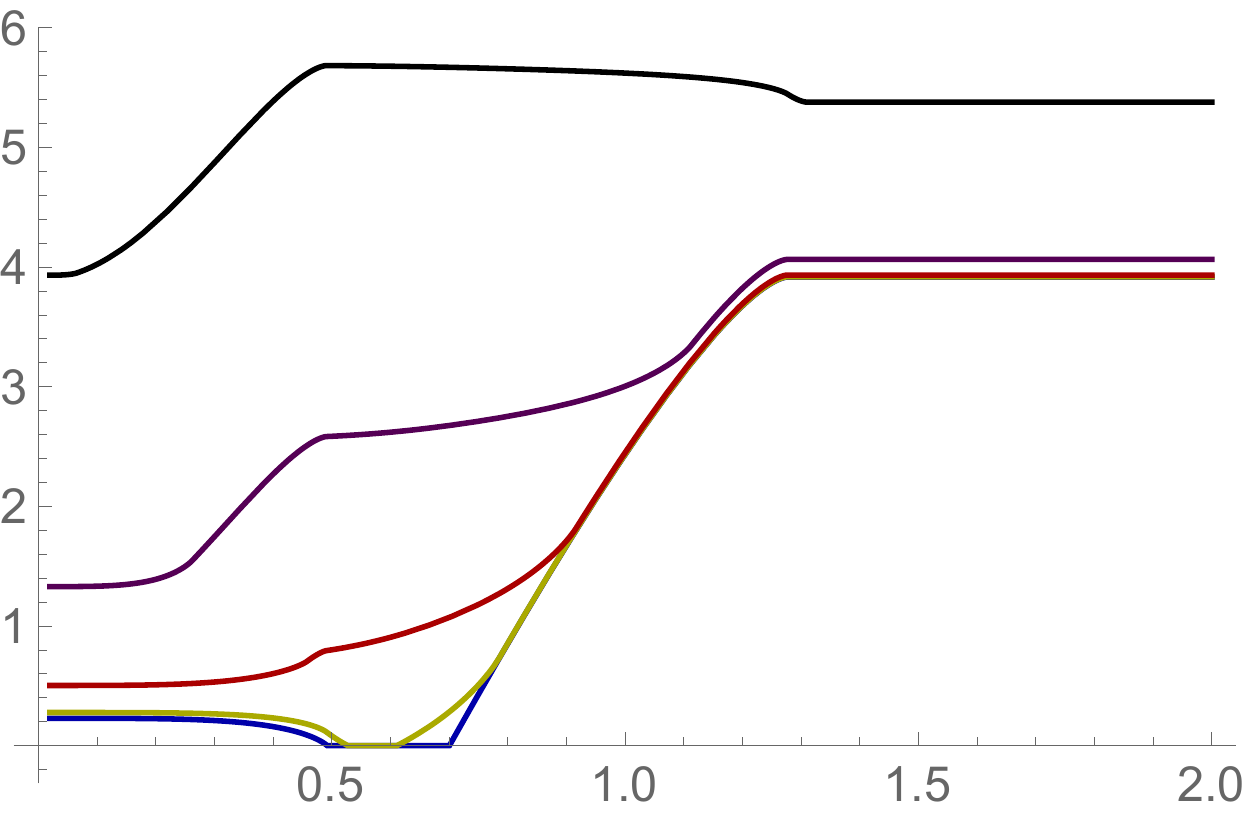}
\begin{picture}(0,0)
\setlength{\unitlength}{1cm}
\put(-7.2,3.9){$I$}
\put(-7.2,8){$I$}
\put(-7.2,12.5){$I$}
\put(-14.6,3.9){$I$}
\put(-14.6,8){$I$}
\put(-14.6,12.5){$I$}
\put(0.2,4.5){$t_\infty$}
\put(0.2,0){$t_\infty$}
\put(0.2,8.8){$t_\infty$}
\put(-8.4,4.4){$t_\infty$}
\put(-8.4,-0.1){$t_\infty$}
\put(-8.4,8.7){$t_\infty$}
\put(-10,3.7){$\boxed{7\pi/16,11\pi/16}$}
\put(-10,7.9){$\boxed{7\pi/16,7\pi/16}$}
\put(-9.6,12.5){$\boxed{\pi/2,\pi/2}$}
\put(-2.7,1.4){$\boxed{5\pi/16,13\pi/16}$}
\put(-2.6,8){$\boxed{9\pi/16,9\pi/16}$}
\put(-2.2,12.5){$\boxed{3\pi/8,5\pi/8}$}
\end{picture}
\caption{\label{fig:hmi} Time evolution of holographic mutual information for $r_+=5$. \textbf{Upper line:} HMI of two intervals of lengths $\pi/2$ and $\pi/2$ (left), and $3\pi/8$ and $5\pi/8$ (right), separated by intervals of length $\pi/24$ (black), $\pi/8$ (purple), $7\pi/24$ (red), $5\pi/12$ (yellow) and $\pi/2$ (blue). \textbf{Middle line (left):} HMI of two intervals of lengths $7\pi/16$, separated by intervals of length $\pi/24$ (black), $\pi/6$ (purple), $7\pi/24$ (red), $\pi/2$ (yellow) and $7\pi/12$ (blue). \textbf{Middle line (right) and lower line:} HMI of two intervals of lengths $9\pi/16$ and $9\pi/16$ (middle line right), $7\pi/16$ and $11\pi/16$ (lower line left), and $5\pi/16$ and $13\pi/16$ (lower line right), separated by intervals of length $\pi/24$ (black), $\pi/6$ (purple), $7\pi/24$ (red), $3\pi/8$ (yellow) and $7\pi/16$ (blue). Note that some of the above lines are always $0$.}
\end{figure}

We observe that the evolution of the HMI strongly resembles the flat boundary case in the upper line and the left panel of the middle line in figure \ref{fig:hmi}. More specifically, considering only same length intervals and varying their separation, the classification of \cite{allaistonni} includes 4 different behaviors:
\begin{itemize}
\item the HMI is $0$ at all times
\item the HMI starts from $0$, becomes positive for some time and ends at $0$
\item the HMI starts from a positive value, forms a ``bump'' and ends at $0$
\item the HMI starts from a positive value, forms a ``bump'' and ends at another positive value, smaller than the initial.
\end{itemize}
Making the two intervals unequal has the effect of ``flattening'' the bump, but no other interesting phenomena arise.\footnote{In \cite{amt} it was shown that the HMI of strip regions in higher dimensions obeys the same classification.}

These 4 cases appear in our study of the compact boundary as well, when the sum of the two intervals is less than half of the total boundary. For instance, the middle left panel depicts the HMI of two equal intervals of length $2\phi_\infty=7\pi/16$, the separation of which increases from top to bottom (for a large enough separation the HMI is always vanishing). This holds until both intervals are of length $\pi/2$ (see the upper left panel). Making the two intervals unequal but keeping the same total length produces a deformation similar to the Poincar\'{e} case (see the upper right panel).

Finite size effects seem to take over when the sum of the lengths of the two boundary intervals exceeds $\pi$. When this sum is large enough, the final value of the HMI is greater than the initial, as can be seen in the middle right panel and the lower line in figure \ref{fig:hmi} (which show the evolution of the HMI for intervals of total length $9\pi/8$, with their separation increasing from top to bottom in each panel). Of course, the initial and final values are dictated by the equilibrium formulae \eqref{eq:adslength} and \eqref{eq:btzlength}, so it is more interesting to observe the interpolation between them.

For two intervals of length $9\pi/16$, we observe that the bump is present when their separation is small enough, but at maximum separation we see a smooth monotonic increase. This is a result of the effective degeneracy of the parameter space due to the symmetry of our configuration, and it holds until one of the complementary intervals has the same length as one of the intervals in question (see the red line in the middle right panel). After this point the bump starts to appear again.

\begin{figure}
\centering
\includegraphics[width=.45\textwidth]{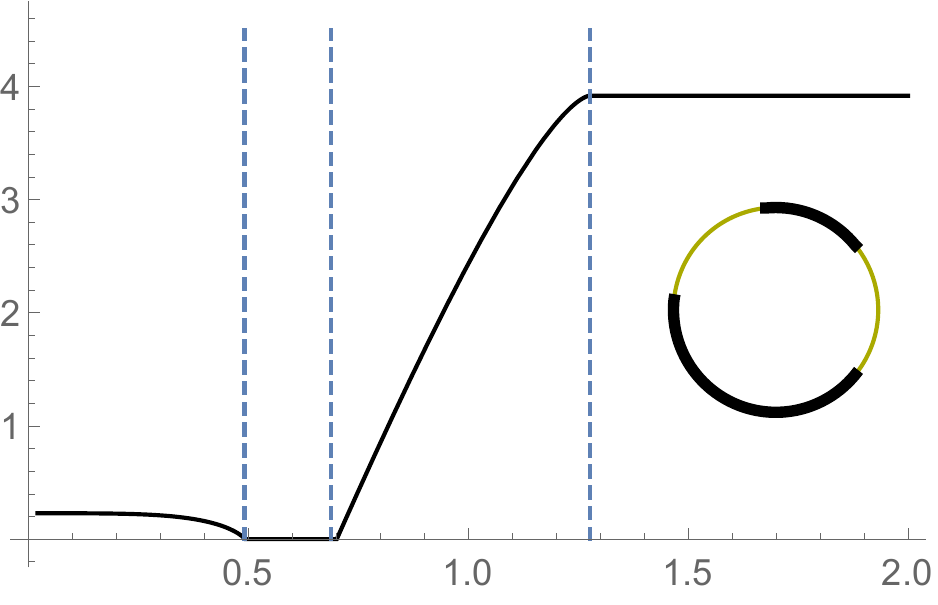}
\hfil
\includegraphics[width=.45\textwidth]{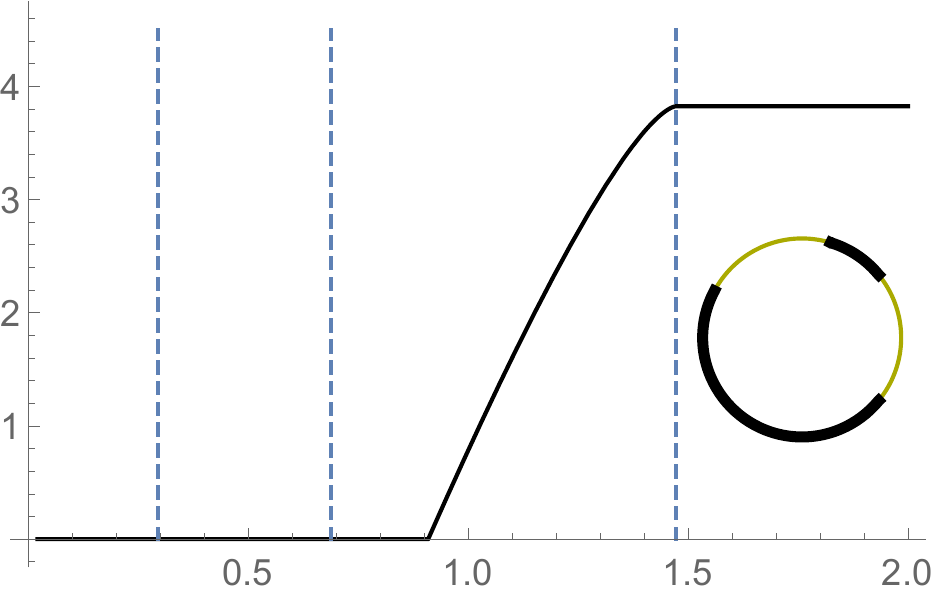}
\hfil
\includegraphics[width=.45\textwidth]{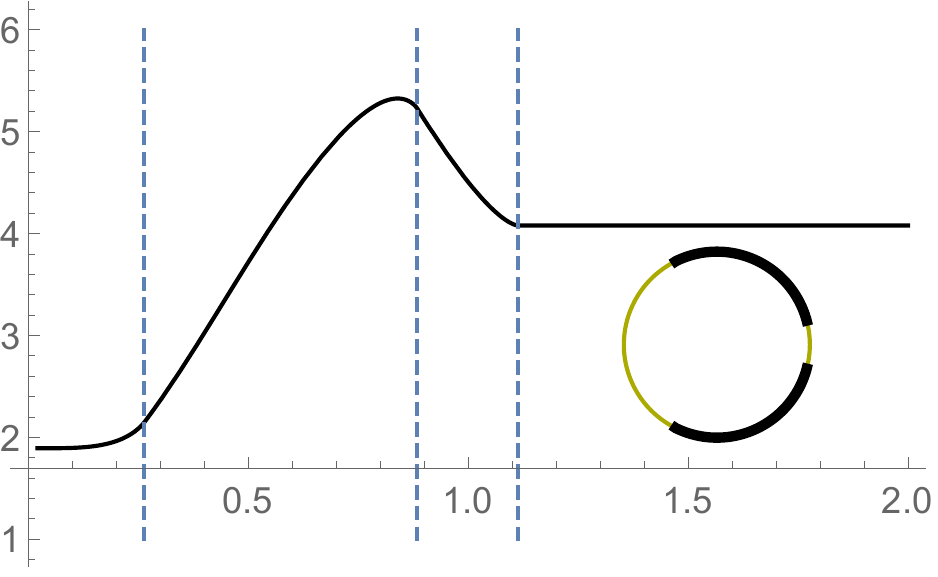}
\hfil
\includegraphics[width=.45\textwidth]{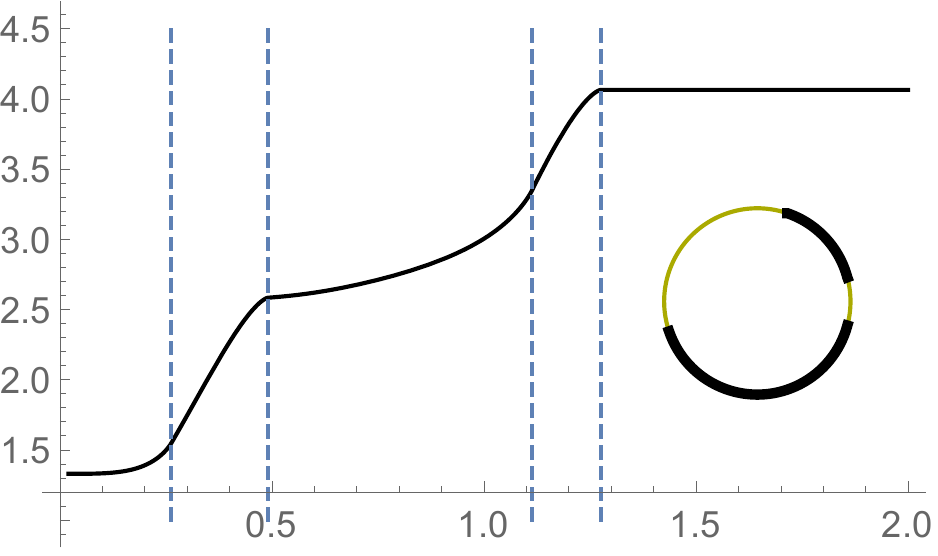}
\begin{picture}(0,0)
\setlength{\unitlength}{1cm}
\put(-7.2,3.8){$I$}
\put(-7.2,8.2){$I$}
\put(-14.6,3.9){$I$}
\put(-14.6,8.2){$I$}
\put(0.2,4.5){$t_\infty$}
\put(0.2,0){$t_\infty$}
\put(-8.1,4.1){$t_\infty$}
\put(-8.1,0.1){$t_\infty$}
\put(-13.4,8.5){$\boxed{(5\pi/16,13\pi/16);(7\pi/16,7\pi/16)}$}
\put(-13.3,3.9){$\boxed{(9\pi/16,9\pi/16);(\pi/6,17\pi/24)}$}
\put(-6,3.9){$\boxed{(5\pi/16,13\pi/16);(\pi/6,17\pi/24)}$}
\put(-6,8.5){$\boxed{(3\pi/16,15\pi/16);(7\pi/16,7\pi/16)}$}
\end{picture}
\caption{\label{fig:hmih} Time evolution of holographic mutual information for $r_+=5$. \textbf{Upper line:} HMI of two intervals of lengths $5\pi/16$ and $13\pi/16$ (left), and $3\pi/16$ and $15\pi/16$ (right), for complementary intervals of lengths $7\pi/16$ in both cases. \textbf{Lower line:} HMI of two intervals of lengths $9\pi/16$ and $9\pi/16$ (left), and $5\pi/16$ and $13\pi/16$ (right) for complementary intervals of length $\pi/6$ and $17\pi/24$ in both cases. In each panel, the intervals in question are graphically represented by the thick black parts of the total boundary circle. The vertical dashed lines mark the saturation times of the different intervals that come into play in computing the HMI (note that in the top line and the bottom left panel two of the dashed lines coincide because, in each case, two of the intervals are of equal length).}
\end{figure}
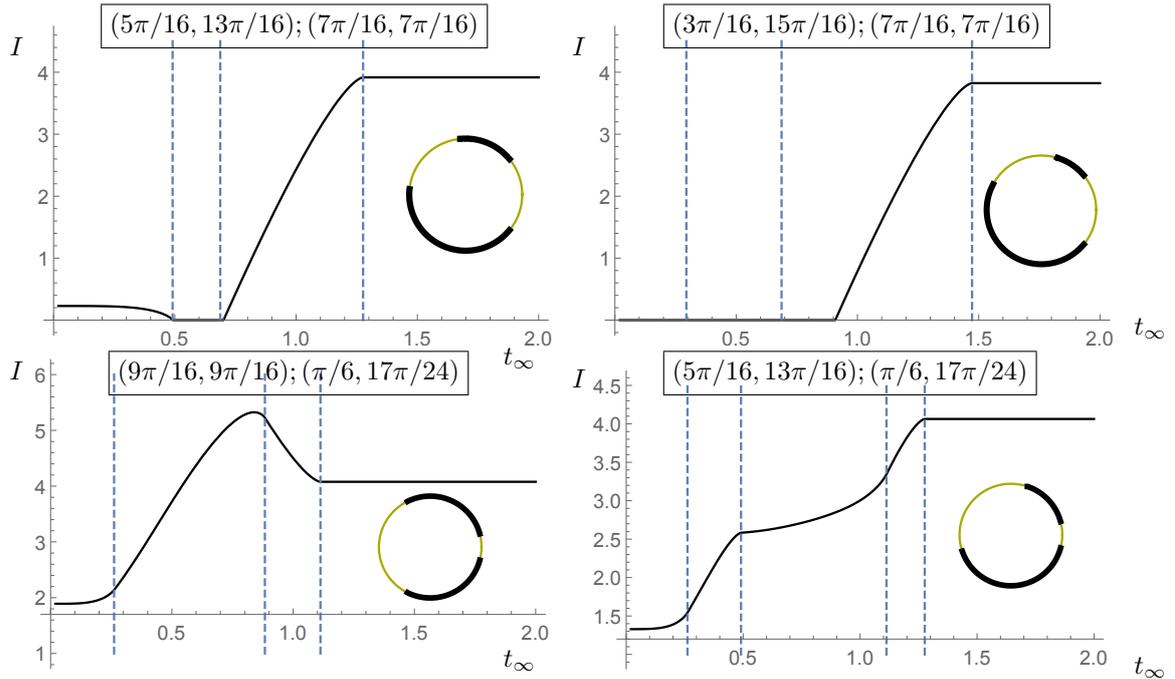

The lower line in figure \ref{fig:hmi} shows how the above considerations are altered when the two intervals have different lengths. In particular, from the lower right panel we can understand that the monotonic behavior appears when one of the complementary intervals becomes equal to one of the intervals in question. When their separation is small, we see a quite deformed bump. When their separation is large, the HMI decreases in the beginning and then increases up to the saturation value. When their separation is large enough, the HMI can even reach $0$ and remain there for some time before it starts increasing. For even more asymmetric intervals, the HMI can begin from $0$ and then start increasing at a specific time\footnote{See bottom right panel of figure \ref{fig:hmih}.}.

Many of the above observations result from the compactness of the boundary. In order to gain a clearer understanding of how this happens, some of the curves of figure \ref{fig:hmi} (plus one not included) are presented in figure \ref{fig:hmih}, together with a graphical representation of the boundary intervals in question.

The behaviors presented in the top line in figure \ref{fig:hmih} can be understood as ``complementary'' to some of the behaviors shown in the middle left panel in figure \ref{fig:hmi} (the purple and the red line, qualitatively). Indeed, having partitioned our system into four intervals, the HMI of two of them equals minus the HMI of the complementary intervals (always bounded by $0$ from below), see eq. \eqref{eq:formi}. More specifically, the intervals in the middle left panel in figure \ref{fig:hmi} are of total length $7\pi/16$ each, whereas in the top line in figure \ref{fig:hmih} \textit{the complementary} intervals are of total length $7\pi/16$ each. These behaviors are absent in the non-compact case, unless we consider infinite size intervals.

The behavior shown in the bottom left panel in figure \ref{fig:hmih} is also a result of the compactness of the boundary. Although a bump is present, the biggest complementary interval is slightly bigger than each of the two intervals in question, and so the HMI decreases a little before saturating at a value larger than its initial one (the fact that only the biggest complementary interval contributes to the decrease can be seen from the vertical dotted lines, which mark the saturation times of the various intervals). In contrast, in the non-compact boundary case, the corresponding interval would be larger than \textit{the sum} of the intervals in question, resulting in a decrease of the HMI to a value always smaller than its initial one.

Now, the evolution shown in the bottom right panel in figure \ref{fig:hmih} is interesting because no bump or ``inverse bump'' is present. As explained above, it can be understood as an intermediate behavior between the cases in which the contribution of one of the intervals in question cancels the contribution of one of the complementary intervals. Two stages of increase are distinguishable in this plot, occurring between the dotted lines for the following reason: the size of the small interval $5\pi/16$ is bigger but comparable to the size of the separation $\pi/6$ and the size of the big interval $13\pi/16$ is slightly bigger than the size of the separation $17\pi/24$. In the beginning of the evolution we can ignore the two biggest intervals because the corresponding geodesics lie mostly in the AdS part of the spacetime and get deformed slowly; thus their contributions almost cancel each other. So, after the smallest of the two complementary intervals has saturated (first dotted line), the main contribution to the HMI comes from the interval of length $5\pi/16$, leading to the first clear stage of increase. In an analogous way we can understand the second stage of increase as the competition between the remaining two intervals. We thus see that this is also a direct result of the compactness of the boundary. Similar arguments can be applied to most of the curves presented in figures \ref{fig:hmi} and \ref{fig:hmih}.

Finally, from figure \ref{fig:hmi} we note that for any configuration, as we bring the two intervals close together, the HMI seems to obtain a fixed shape, which is then only shifted to larger values. This happens because, to first order, the HEE of a small interval is just a (negative) constant (see equations \eqref{eq:adslength}, \eqref{eq:btzlength} and also figure \ref{fig:hee}).

\subsection{Comparison with adiabatic approximation}

Finally, let us compare these results with what we would get using the adiabatic approximation. From the boundary theory point of view, we slowly inject energy to the system, so that it (approximately) stays in equilibrium during the whole evolution. Holographically, this amounts to effectively making the infalling shell thick enough, so that spacetime looks like a BTZ black hole with its radius slowly increasing in time. In other words, when the mass function $m(v)$ in eq. \eqref{eq:f} is slowly increasing\footnote{Note that the null energy condition does not allow $m(v)$ to decrease.}, we can approximate our fixed-$v$ spacetime slice with a BTZ spacetime slice, where $r_+=\sqrt{m(v)}$. The spacelike geodesics giving the HEE will then have their equilibrium forms, i.e. eq. \eqref{eq:btzlength}, with the replacement $r_+=\sqrt{m(v)}$. This is equivalent to forcing our system to remain in equilibrium. In the rest of this subsection, $r_+$ will refer to the final black hole radius $r_+=\lim_{v\rightarrow\infty}\sqrt{m(v)}$ and we will use $m(v)$ to describe the intermediate stages of the evolution.

Pure AdS$_3$ is obtained for $m(v)=-1$. As $m(v)$ increases, a naked conical singularity forms until $m(v)=0$, while for $m(v)>0$ an event horizon appears\footnote{The Hawking-Page transition for $0<m(v)<1$ does not concern us here.}. It is natural to consider the mass profile
\begin{equation}\label{eq:mtanh}
m(v)=\frac{r_+^2+1}{2}\tanh\left(\frac{v}{v_t}\right)+\frac{r_+^2-1}{2},
\end{equation}
smoothly interpolating between $m=-1$ and $m=r_+^2$, with $v_t$ corresponding to the effective ``thickness'' of the shell. The limit $v_t\rightarrow0$ reproduces the previous analysis of the instantaneous quench.

In \cite{hubenyrangamanitakayanagi} the adiabatic approximation was checked against numerical calculations for $r_+=1$ and $v_t=1$, and a very good agreement was observed for the HEE of intervals of various sizes. This implies that we can compute the HMI using the adiabatic approximation with the mass profile $m(v)=\tanh(v)$ and we are guaranteed to obtain trustworthy results for a slow enough quench. So we are going to consider this case, with our analysis also persisting to other values of $r_+$.

\begin{figure}
\centering
\includegraphics[width=.50\textwidth]{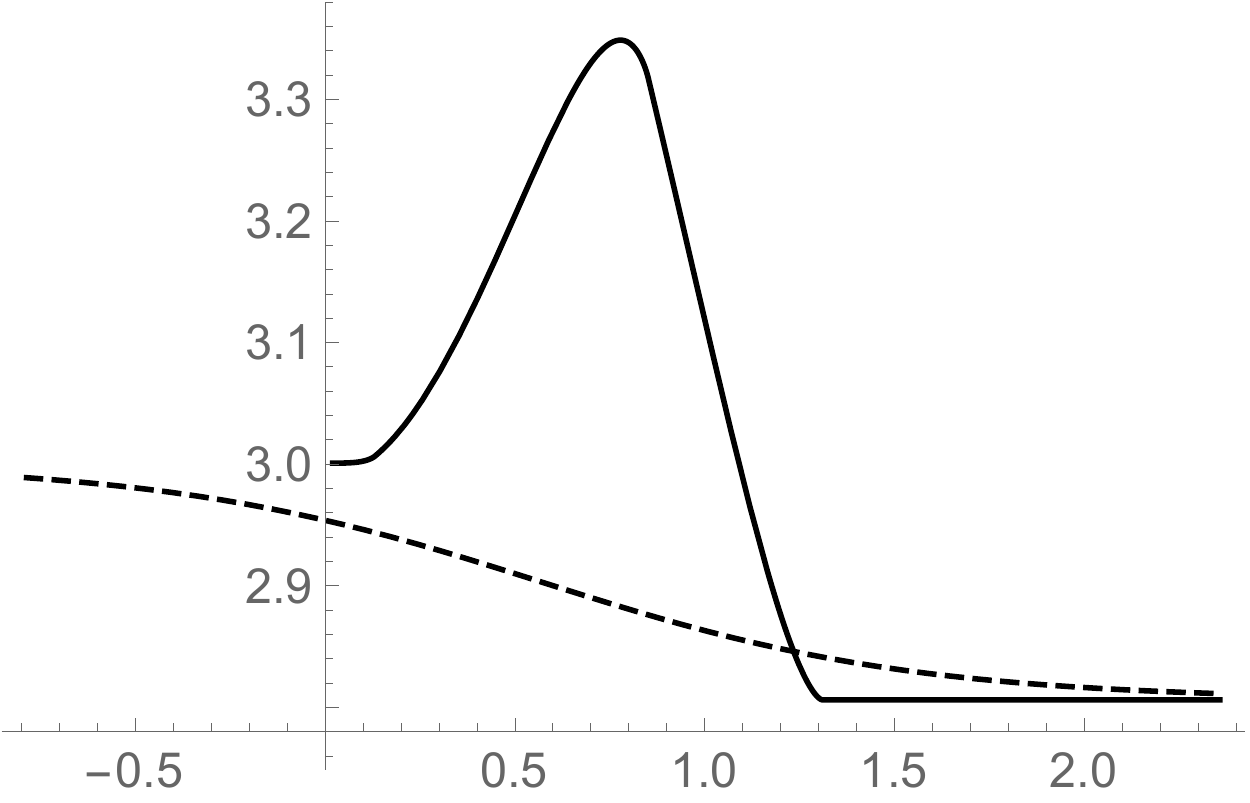}
\begin{picture}(0,0)
\setlength{\unitlength}{1cm}
\put(-6.5,4.7){$I$}
\put(0,0){$t_\infty$}
\put(-2.5,4){$\boxed{13\pi/24,13\pi/24}$}
\end{picture}
\caption{\label{fig:adhmi1} Time evolution of holographic mutual information among two intervals of lengths $13\pi/24$ separated by an interval of length $2\pi/24$, for $r_+=1$. The continuous curve represents the case of an infinitely thin shell, while the dashed curve represents the case of a shell with effective thickness $v_t=1$.}
\end{figure}
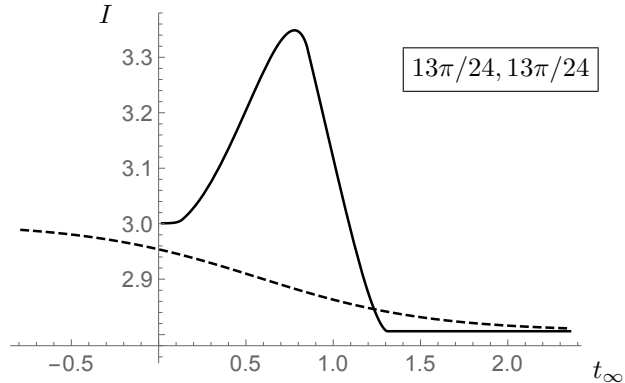

Figure \ref{fig:adhmi1} confirms the expectations outlined in the introduction, in the specific case of two intervals of lengths $13\pi/24$ separated by an interval of length $20\pi/24$: even though the infinitely thin shell produces a ``bump'' in the HMI, the thick shell predicts a smooth monotonic evolution. A similar monotonic evolution between the initial and final value of the HMI was also confirmed for many other choices of intervals, in contrast to the variety of different behaviors presented in figure \ref{fig:hmi}.

\begin{figure}
\centering
\includegraphics[width=.40\textwidth]{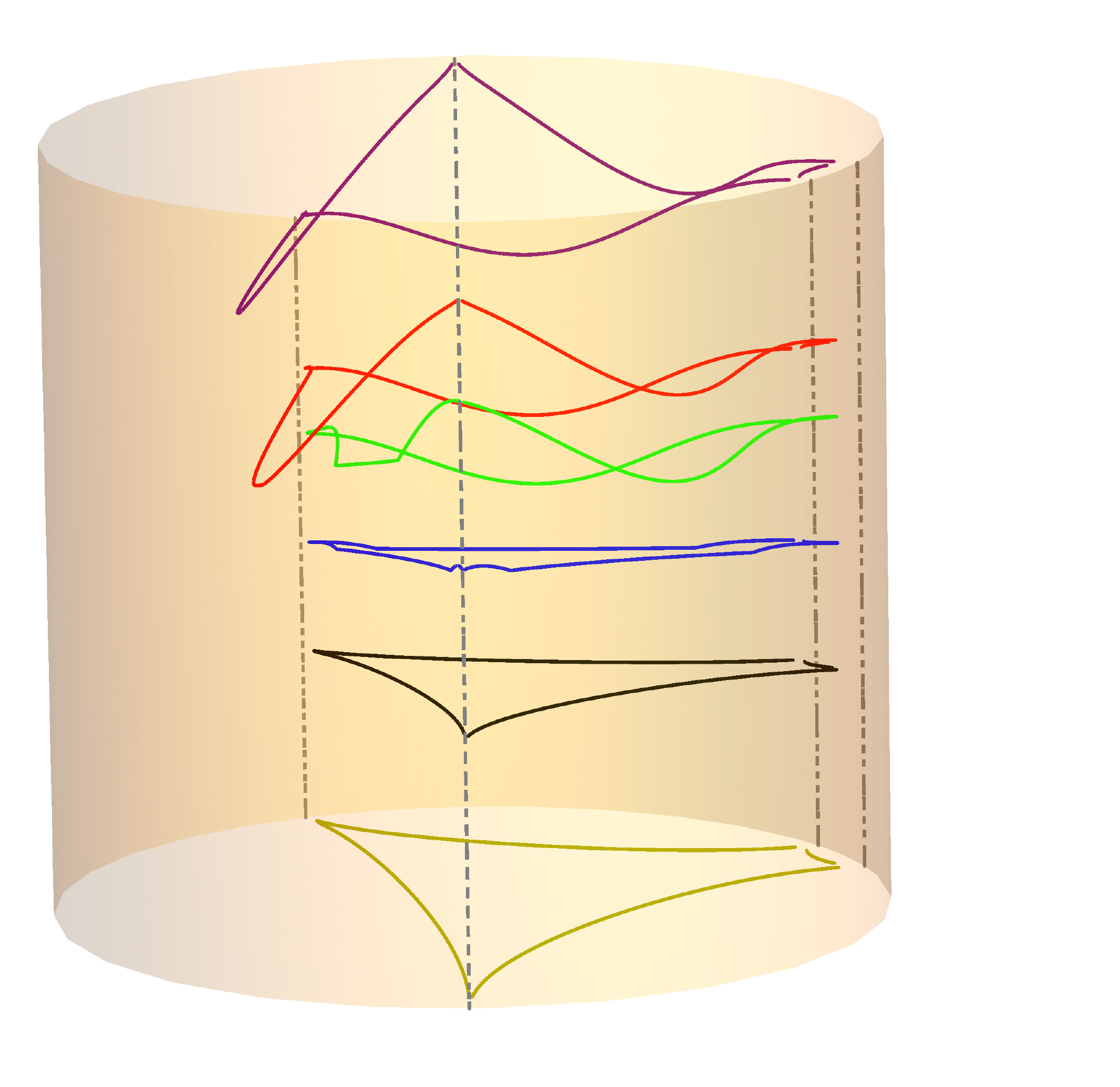}
\hfil
\includegraphics[width=.40\textwidth]{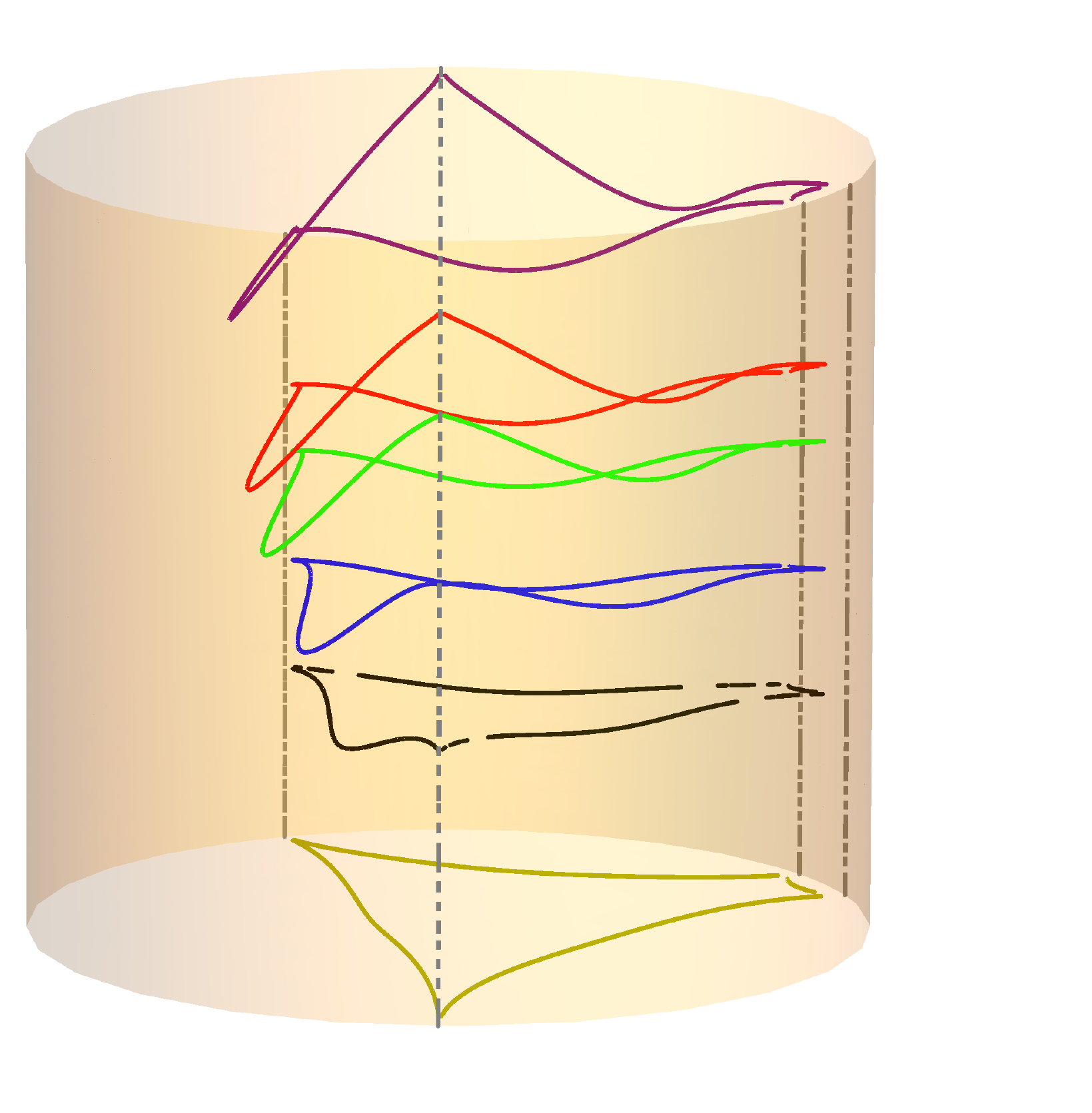}
\hfil
\includegraphics[width=.45\textwidth]{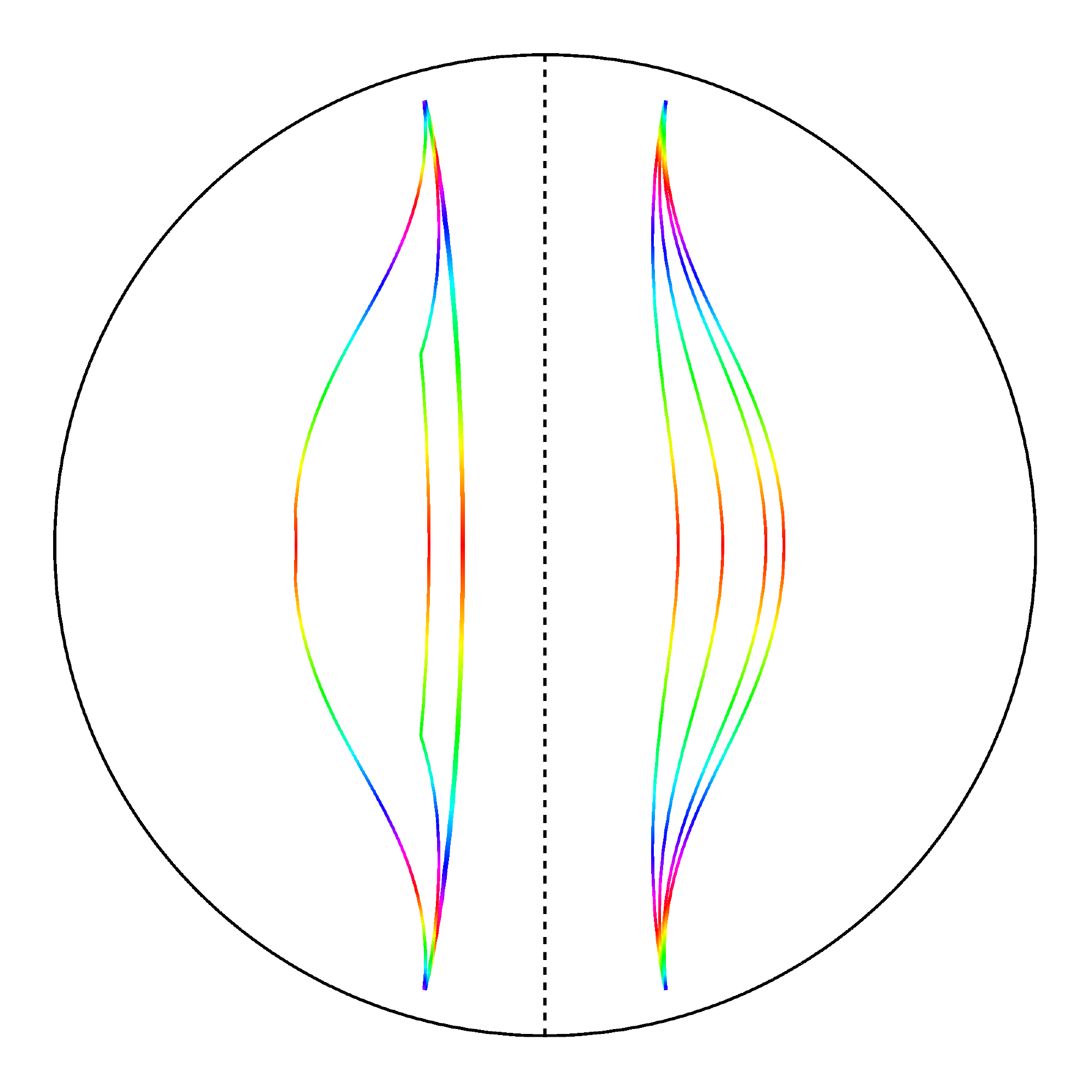}
\caption{\label{fig:col1} \textbf{Upper line}: The conventions are as in figure \ref{fig:1112plot2}. Here, $t_\infty$ runs from $-\pi/4$ up to 2 and the black hole radius is $r_+=1$. Neither the shell nor the black hole horizon is depicted. The boundary is partitioned in 4 intervals of lengths $20\pi/24$, $13\pi/24$, $13\pi/24$ and $4\pi/24$, as indicated by the dashed lines. The corresponding spacelike geodesics are presented for $t_\infty=-\pi/4$, 0, 0.5, 1, 1.3 and 2, for the non-adiabatic (left) and the adiabatic case (right). Note that the black geodesics seem to be discontinuous only due to numerical inaccuracies. \textbf{Lower line}: Combination of $\rho-\varphi$ geodesic projections in the non-adiabatic (left part) and the adiabatic case (right part) separated by the dashed line, for an interval of length $20\pi/24$ at times $t_\infty=0$, 0.5, 1 and 1.3. The color coding indicates the rate at which the affine parameter $s$ changes along the geodesics: the quicker the color changes, the faster $s$ increases. More specifically, $s$ increases by 2 as we go from red back to red. Note that the geodesics get deformed in an outwards direction as $t_\infty$ increases, and that the first two geodesics in the non-adiabatic case almost coincide. In all of the above plots, we have imposed a cut-off for the geodesics at $r=10$.}
\end{figure}

In order to highlight the differences between the adiabatic and the non-adiabatic evolution, both cases are presented in figure \ref{fig:col1}. In particular, the top left panel shows all of the geodesics that come into play in the non-adiabatic evolution, from $t_\infty=-\pi/4$ up to $t_\infty=2$ (for clarity, the shell is not depicted). The top right panel shows the analogous process using the adiabatic approximation. In the bottom panel we can see the projection of the four intermediate geodesics on the $\rho-\varphi$ plane, with the color coding indicating the rate of change of the affine parameter $s$ along the geodesic. We have imposed a cut-off for the geodesics at $r=10$.

In figure \ref{fig:col1} we can observe the geometric differences between the two processes, the sharpest being the way the geodesics are ``deformed'': in the non-adiabatic evolution we see that the ``perturbations'' propagate from the endpoints towards the middle part, whereas in the adiabatic evolution the singularity ``pushes'' the geodesics mainly at their middle part. This produces a tension in the following sense: big intervals feel the effect of the singularity formation early in their adiabatic evolution, since they reach deep inside the bulk geometry. In contrast, in the non-adiabatic case, the length increase is slow in the beginning because the deformation starts from the UV (i.e. the bulk asymptotic AdS region). The length increases significantly only when the shell has ``collapsed enough'' so that the geodesic can feel the effect of the mass. This can be seen in figure \ref{fig:col1} (lower line), taking into account the fact that the next to last geodesics in both cases have the same length. In the non-adiabatic case we observe that the gravitational pull of the shell makes the geodesic long in the region close to the shell, compensating for the middle part which is in pure AdS. Similar arguments apply to small intervals, too: in the instantaneous quench case the shell passes through the geodesics in the beginning of the collapse, and so they quickly saturate to their equilibrium form. In the adiabatic case, it takes them a long time to ``feel'' the effects of the singularity, since they are close to the boundary\footnote{We should mention that the recent work \cite{abgh} gave strong evidence that the entanglement scrambles maximally in the holographic systems considered by the authors, and so the bump in the HMI is only a result of the different saturation times of different intervals. Since the bump disappears in the adiabatic approximation in the non-compact case as well, it can be inferred that this is what produces the tension between the fast and the slow quench, as was also explained from a geometrical point of view above.}.

It is important to note that making the parameter $v_t$ smaller while still employing the adiabatic approximation does not result in the HMI interpolating between the behaviors presented in figure \ref{fig:adhmi1}. This shows that the instantaneous quench cannot be modelled by an adiabatic process, as expected from the above discussions. However, the fact that in both cases there is only one family of geodesics which gives the HHE of a boundary interval suggests that this should also hold in the numerical solutions of the cases with $0<v_t<1$. We expect that, as $v_t$ grows from $0$ to $1$, the cusps at the places where a geodesic crosses the infinitely thin shell get ``smoothed out'' gradually, until they look similar to the geodesics in the adiabatic approximation. Accordingly, we expect the HMI to interpolate between the dashed and the continuous curves in figure \ref{fig:adhmi1}, with the bump starting to form at some intermediate stage.

\section{Conclusion and discussion}

To sum up, in this paper we have explored the evolution of the holographic entanglement entropy and the mutual information after a quantum quench with the dual description of Vaidya-BTZ spacetime. Figure \ref{fig:hmi} presents a classification of the qualitatively different behaviors of the HMI, confirming the behaviors that can be intuitively understood by the arguments in subsection \ref{subsection:hmi}. We also highlighted the geometric differences between the instantaneous and the slow quench.

We finish our work with a few comments and open questions. Firstly, it is interesting to note that the recent proposal of \cite{leichenauermoosa} for calculating the (vacuum subtracted) entanglement entropy using the ``entanglement tsunami'' picture of \cite{liusuh1},\cite{liusuh2} seems to be able to capture many qualitative features of the HMI even in our compact boundary case. Although some modifications are required in order to achieve good agreement (it only describes the linear part of the HEE in the non-compact case, which is absent here), it provides strong support for the usefulness of an ``entanglement tsunami'' toy model. The importance of such a description was also hinted in our comparison of the thin shell and the thick shell spacetimes: we saw that the ``deformation'' of the geodesics propagated from the endpoints and the midpoint respectively, resulting in the different evolutions of the HMI. Now, we expect that a slightly thick shell will give rise to a tsunami with a slightly thick wavefront. However, this picture should break down as the thickness of the wavefronts becomes comparable to the size of the intervals in question (which are even comparable to the whole system in the compact case), and indeed the behavior observed in the adiabatic approximation was quite different from what a tsunami picture would suggest.

Finally, we should mention that there are many generalizations of the simple Vaidya-BTZ model that one could consider. Adding charge and angular momentum to the black hole would be the most natural generalization, with the geodesic structure of the spacetime becoming complicated enough to allow for novel phenomena in the time evolution of the HMI. More importantly, a systematic study of higher dimensional Vaidya-type spacetimes in global coordinates, along the lines of \cite{liusuh2}, would be highly desirable. This would illuminate the possibility of observing genuinely finite size effects, invisible to the latter analysis. Of course, a direct calculation of these processes with CFT techniques, analogous to those recently developed for the non-compact case (see for example \cite{hartman}, \cite{faulkner}), would allow us to obtain a spherical perspective on thermalization.

\acknowledgments

It is a pleasure to thank Veronika Hubeny for her guidance throughout the different stages of this project and Henry Maxfield for useful comments on the manuscript. I acknowledge support by a Faculty of Science Durham Doctoral Scholarship.

\appendix
\section{Solving the geodesic equations}\label{appendixa}

Here we present the solution of the geodesic equations \eqref{eq:geodeq}. First they must be solved inside the shell (i.e. for $v<0$), then outside (i.e. for $v>0$), and finally the solutions must be patched together on the shell $v=0$, taking care to keep only those that end on the boundary. Most of the following results were also obtained in \cite{hubenyrangamanitonnitonni}.

\paragraph{Inside the shell:}
Recalling that we only consider symmetric geodesics, the energy should vanish inside the shell, $E=0$, and we can set the affine parameter $s$ to be equal to $0$ at the minimum $v$ point $v_{min}$. Then, for $v<0$, the geodesic equations \eqref{eq:geodeq} become
\begin{subequations}\label{eq:geodeqinside}
\begin{align}
L & =(r^i)^2\dot{\varphi}^i,\label{eq:geodeqinside:1}\\
\dot{v}^i & =\frac{\dot{r}^i}{(r^i)^2+1},\label{eq:geodeqinside:2}\\
(\dot{r}^i)^2 & = -\left(\frac{L^2}{(r^i)^2}-1\right) ((r^i)^2+1).\label{eq:geodeqinside:3}
\end{align}
\end{subequations}
The solution of \eqref{eq:geodeqinside:3} subject to $r(s=0)=L$ is
\begin{equation}\label{eq:rin}
r^i(s,L)=\sqrt{\frac{(L^2+1)\cosh(2s)+L^2-1}{2}}
\end{equation}
and the solution of \eqref{eq:geodeqinside:2} subject to the same condition is
\begin{equation}\label{eq:vin}
v^i(s,L,v_{min})=\tan^{-1}\left(\frac{r^i(s,L)-L}{1+r^i(s,L)L}\right)+v_{min}.
\end{equation}
Without loss of generality we can set $\varphi(s=0)=0$, and then \eqref{eq:geodeqinside:1} implies that
\begin{equation}\label{eq:phiin}
\varphi^i(s,L)=\tan^{-1}\left(\frac{\tanh(s)}{L}\right).
\end{equation}
The parameter value $s_0$ at which the geodesic encounters the shell $v=0$, will actually give us (half of) the length of the AdS part $v<0$. Setting $r^i(s_0,L)\equiv r_s$, from \eqref{eq:vin} evaluated at $s_0$ we get:
\begin{equation}\label{eq:vminlrs}
v_{min}=\tan^{-1}\left(\frac{L-r_s}{1+r_sL}\right).
\end{equation}
We can now invert \eqref{eq:rin} and \eqref{eq:phiin} to find (half of) the proper length inside the shell $\ell^i\equiv s_0$ and $\varphi|_{shell}$ as functions of the two parameters $r_s$ and $L$:
\begin{equation}\label{eq:s0in}
s_0=\frac{1}{2}\log\left(\frac{2r_s^2-L^2+1+2\sqrt{(r_s^2+1)(r_s^2-L^2)}}{L^2+1}\right)
\end{equation}
and
\begin{equation}\label{eq:phishell}
\varphi|_{\text{shell}}=\tan^{-1}\left(\frac{1}{L}\sqrt{\frac{r_s^2-L^2}{r_s^2+1}}\right).
\end{equation}\\

\paragraph{Outside the shell:}
In region $v>0$, the geodesic equations take the form
\begin{subequations}\label{eq:geodeqoutside}
\begin{align}
L & =r^2\dot{\varphi}^o,\label{eq:geodeqoutside:1}\\
E & =((r^o)^2-r_+^2)\dot{v}^o-\dot{r}^o,\label{eq:geodeqoutside:2}\\
(\dot{r}^o)^2 & = E^2 -\left(\frac{L^2}{(r^o)^2}-1\right)((r^o)^2-r_+^2).\label{eq:geodeqoutside:3}
\end{align}
\end{subequations}
The relation $E=-\frac{(r_+^2+1)}{2}\dot{v}|_{v=0}$ \eqref{eq:deltae}, the continuity of $\dot{v}$ across the shell and the explicit expression for $v^i$, eq. \eqref{eq:vin}, give the energy in the region $v>0$
\begin{equation}\label{eq:energyl}
E=-\frac{(r_+^2+1)}{2r_s}\sqrt{\frac{r_s^2-L^2}{r_s^2+1}}.
\end{equation}
Also, evaluating eq. \eqref{eq:geodeqoutside:2} at $v=0^+$, we find
\begin{equation}\label{eq:rdotout}
\dot{r}^o|_{\text{shell}}=-\frac{E(2r_s^2-r_+^2+1)}{(r_+^2+1)} \equiv \beta.
\end{equation}
We observe that the velocity of the geodesic just after it crosses the shell is positive when $r_s^2>(r_+^2-1)/2$ and negative when $r_s^2\leq(r_+^2-1)/2$.
Setting now $s=0$ at the point where the geodesic crosses the shell, i.e. $r^o(s=0)=r_s$ and $v^o(s=0)=0$, we can solve \eqref{eq:geodeqoutside:3} with the aid of eq. \eqref{eq:energyl}, to get
\begin{equation}\label{eq:rout}
r^{o}(s,\alpha,\beta,r_+,r_s)= \frac{1}{2}\sqrt{-2\alpha+2(2r_s^2+\alpha)\cosh(2s)+4\beta r_s\sinh(2s)},
\end{equation}
where we set $\alpha=E^2-L^2-r_+^2$ for brevity.

Eq. \eqref{eq:geodeqoutside:3} determines the behavior of $(\dot{r}^o)^2$, and we can thus find the necessary and sufficient conditions which the geodesics should obey in order to end on the boundary:
\begin{itemize}
  \item If there is no turning point, i.e. $(\dot{r}^o)^2>0$ always, the geodesics should emerge from the shell with a positive velocity, i.e. $\dot{r}^o(s=0)>0$. Eq. \eqref{eq:geodeqoutside:3} implies that the minimum of $(\dot{r}^o)^2$ is at $\sqrt{L r_+}$, taking the value $E^2-(L-r_+)^2$ there. This means that when $E^2>(L-r_+)^2$, we necessarily have $r_s^2\geq(r_+^2-1)/2$.
  \item If there is a turning point, i.e. $(\dot{r}^o)^2=0$ at some point, this should be at a value $\tilde{r}$ of $r$ with $\tilde{r}\leq r_s$, so that the geodesics emerging from the shell with positive velocity don't get affected, while the ones emerging with negative velocity do change their direction. In this case eq. \eqref{eq:geodeqoutside:3} implies that $\tilde{r}\equiv\sqrt{-\alpha+\sqrt{\alpha^2-4(L r_+)^2}} / \sqrt{2}\leq r_s$.
\end{itemize}

Keeping these conditions in mind, we can proceed to solve eq. \eqref{eq:geodeqoutside:1}:
\begin{equation}\label{eq:phiout}
\varphi^{o}(s,\beta,L,r_+,r_s,\varphi_s)=\frac{1}{r_+}\tanh^{-1}\left( \frac{L r_+ \tanh(s)}{r_s(r_s+\beta\tanh(s))} \right) +\varphi|_{\text{shell}}.
\end{equation}
Similarly, the solution of eq. \eqref{eq:geodeqoutside:2} subject to $v^o(s=0)=0$ is
\begin{equation}\label{eq:vout}
v^{o}(s,\beta,E,r_+,r_s) = \frac{1}{r_+} \left[ \tanh^{-1}\left( \frac{E r_+\tanh(s)}{-r_+^2+r_s(r_s+\beta\tanh(s))} \right) + \tanh^{-1}\left( \frac{r_+(r_s-r^{o}(s))}{r_+^2-r_sr^{o}(s)}\right) \right].
\end{equation}

We can now take the limit $s\rightarrow\infty$ in \eqref{eq:vout} and solve for the energy $E$ in terms of $r_+$, $r_s$ and the time on the boundary $t_\infty\equiv v^o(s\rightarrow\infty)$. Explicitly,\\
\begin{equation}\label{eq:energyt}
E = \frac{(r_+^2+1)(r_+-r_s T)}{T(2r_s^2+r_+^2+1)-2r_+r_s},
\end{equation}
where $T\equiv\tanh(r_+t_\infty)$. So, through eq. \eqref{eq:energyl} we can express $L$ as a function of $r_+$, $r_s$ and $t_\infty$, as well as the quantities $s_0$ and $\varphi|_{\text{shell}}$. (Half of) the extent of the boundary interval will be given by the $s\rightarrow\infty$ limit of \eqref{eq:phiout}, while (half of) the proper length in the BTZ part can be found by inverting eq. \eqref{eq:rout}. Specifically, assuming $r_\infty$ to be a large radial cut-off, we find
\begin{equation}\label{eq:sout}
\exp(2s_\infty) = \frac{4r_\infty^2}{2r_s^2+\alpha+2\beta r_s}+\mathcal{O}(r_\infty^0).
\end{equation}
Expressing all of the above as functions of $r_+$, $r_s$ and $t_\infty$, we are led to the results \eqref{eq:phitotal} and \eqref{eq:lengthtotal} presented in section \ref{geosol}. In an analogous way, the conditions for the geodesics to end on the boundary boil down to eq. \eqref{eq:constraints}.

\section{Early and late time evolution of holographic entanglement entropy}\label{appendixb}

Here we present the calculations of subsection \ref{subsection:hee}.

Firstly, it is convenient to make the change of variables $r_s\leftrightarrow a$ defined by:
\begin{equation}\label{rsadef}
r_s=\frac{r_+^2-1+(r_+^2+1)\sqrt{1-\tanh(r_+t_\infty)^2}}{2r_+\tanh(r_+t_\infty)}(1-a)+a\frac{r_+}{\tanh(r_+t_\infty)}.
\end{equation}
This way $a$ interpolates between $\nu_1$ for $a=0$ and $\nu_2$ for $a=1$, while it simplifies the constraints \eqref{eq:constraints} to
\begin{equation}\label{eq:a01}
0\leq a\leq1.
\end{equation}

So, we can express $\varphi_\infty$ and $\ell$ as functions of $a$ in the following way:
\begin{equation}\label{eq:phia}
\begin{aligned}
\varphi_\infty(r_+,a,t_\infty) & =\tan^{-1}\left(\frac{\sqrt{2}(1-a)\sinh(r_+ t_\infty/2)} {r_+\sqrt{a(2-a+a\cosh(r_+ t_\infty))}}\right)\\
& +\frac{1}{r_+}\text{tanh}^{-1}\left(\frac{\sqrt{2a(2-a+a\cosh(r_+ t_\infty))}\sinh(r_+ t_\infty/2)} {1-a+a\cosh(r_+ t_\infty)}\right)
\end{aligned}
\end{equation}
and
\begin{equation}\label{eq:lengthtotala}
\ell(r_+,a,t_\infty)=\ln \left( \frac{((-3(a-1)^2+(-3a^2+4a+1)r_+^2)\sinh\left(\frac{r_+ t_\infty}{2}\right) + ((a-1)^2+(a^2+1)r_+^2)\sinh\left(\frac{3r_+ t_\infty}{2}\right))^2}{8r_+^4(-1+(2-a)a(r_+^2+1)+(1+a(-2+a+a r_+^2))\cosh(r_+ t_\infty))} \right).
\end{equation}

Fixing $r_+$, we are looking for paths $a(t_\infty)$ such that $\varphi_\infty$ in eq. \eqref{eq:phia} is constant. So, forcing the time derivative of eq. \eqref{eq:phia} to vanish, we get the relation
\begin{equation}\label{eq:ader}
a'=\frac{a r_+\coth\left(\frac{r_+ t_\infty}{2}\right) (r_+^2-1+\cosh(r_+ t_\infty)+a(2+r_+^2-2\cosh(r_+ t_\infty))+2a^2(r_+^2+1)\sinh\left(\frac{r_+ t_\infty}{2}\right)^2)} {(a-1)(r_+^2+1-\cosh(r_+ t_\infty)+2a(r_+^2+1)\sinh\left(\frac{r_+ t_\infty}{2}\right)^2)},
\end{equation}
where we denote by prime the derivative with respect to $t_\infty$. Using \eqref{eq:ader} in the time derivative of eq. \eqref{eq:lengthtotala} we find that along paths of constant $\varphi_\infty$
\begin{equation}\label{eq:lengthdera}
\ell'|_{\varphi_\infty}=\frac{2r_+(r_+^2+1)(1-a)\sinh(r_+ t_\infty)} {r_+^2-1+(r_+^2+1)(2-a)a+\cosh(r_+ t_\infty)+(r_+^2+a(-2+a(r_+^2+1)))\cosh(r_+ t_\infty)}.
\end{equation}

Now, recalling the restriction \eqref{eq:a01}, it can be shown that $\ell'|_{\varphi_\infty}\geq0$ for any $t_\infty>0$, with the equality holding only when $t=0$ or $a=1$, i.e. when the geodesics start evolving and when they reach their final BTZ form. This shows in particular that the time derivative of the HEE is continuous.

We can see in more detail the early and late time behavior of the HEE by expanding perturbatively $a(t_\infty)$ around the points $t_\infty=0$ and $t_\infty=\varphi_\infty$. Specifically, we see that if
\begin{equation}\label{eq:earlya}
a(t_\infty)=\frac{1}{4\tan\phi}t_\infty^2+\frac{6+(12+r_+^2)\tan\phi^2}{48\tan\phi^4}t_\infty^4 + \mathcal{O}(t_\infty^5),
\end{equation}
then
\begin{equation}\label{eq:earlyphia}
\varphi_\infty(t_\infty)=\phi + \mathcal{O}(t_\infty^3)
\end{equation}
and substituting eq. \eqref{eq:earlya} in the length \eqref{eq:lengthtotala} and expanding in $t_\infty$, we find
\begin{equation}\label{eq:earlytimehee}
\ell=2\ln(\sin\phi)+\frac{r_+^2+1}{2}t_\infty^2+\mathcal{O}(t_\infty^3),
\end{equation}
as was also stated in subsection \ref{subsection:hee}.

Similarly, expanding $a(t_\infty)$ in $\phi-t_\infty$ we find
\begin{equation}\label{eq:latea}
\begin{aligned}
a(t_\infty)=1 & - \frac{\sinh(r_+\phi)\sqrt{2r_+\tanh(r_+\phi)}}{\cosh(r_+\phi)-1} \sqrt{(\phi-t_\infty)}\\
& - \frac{2\sinh(r_+\phi)(-3r_+^2+(r_+^2+1)\tanh(r_+\phi))}{3r_+\cosh(r_+\phi)-1}(\phi-t_\infty) + \mathcal{O}((\phi-t_\infty)^{3/2}),
\end{aligned}
\end{equation}
so that
\begin{equation}\label{eq:latephia}
\varphi_\infty(t_\infty)=\phi + \mathcal{O}((\phi-t_\infty)^2).
\end{equation}
Substituting in \eqref{eq:lengthtotala} and expanding in $\phi-t_\infty$ we find
\begin{equation}\label{eq:latetimehee}
\ell=2\ln\left(\frac{1}{r_+}\sinh(r_+\phi)\right)-\frac{2\sqrt{2}(r_+^2+1)\sqrt{\tanh(r_+\phi)}}{3\sqrt{r_+}} (\phi-t_\infty)^{3/2}
+\mathcal{O}((\phi-t_\infty)^2).
\end{equation}

An analogous analysis can be performed in the Poincar\'{e} case, resulting in similar expressions with some slight changes in the coefficients of the above formulas. They can be computed exactly by taking the Poincar\'{e} limit \eqref{poincarescaling}), reproducing the Vaidya-BTZ case of the general results of \cite{liusuh2} in the early growth and saturation regimes.

\section{Rate of growth in Poincar\'{e} Vaidya-BTZ}\label{appendixc}

Taking the Poincar\'{e} limit of the corresponding expressions in \ref{appendixb}, we obtain the time derivative of the length along paths of constant $\varphi_\infty$
\begin{equation}\label{eq:lengthderapoi}
\ell'^{P}|_{\varphi_\infty}=\frac{2r_+(1-a)\sinh(r_+ t_\infty)} {(2-a)a+1+(1+a^2)\cosh(r_+ t_\infty)},
\end{equation}
where now
\begin{equation}\label{rsadefpoi}
r_s=r_+\frac{1+\sqrt{1-\tanh(r_+t_\infty)^2}}{2\tanh(r_+t_\infty)}(1-a)+a\frac{r_+}{\tanh(r_+t_\infty)}
\end{equation}
and so $a$ is still bounded by $0$ and $1$. Now, since the derivative of eq. \eqref{eq:lengthderapoi} with respect to $a$ at constant $t_\infty$ is always negative, we find that $\ell'^{P}|_{\varphi_\infty}$ is a decreasing function of $a$. Thus, at every $t_\infty$ we have
\begin{equation}
\mathfrak{R}^{(2)P}(t_\infty)=\frac{1}{2r_+}\ell'^{P}\leq\frac{1}{2r_+}\ell'^{P}|_{a=0}=\tanh{\left(\frac{r_+t_\infty}{2}\right)}<1.
\end{equation}
We conclude that, in the flat boundary case, the rate of growth $\mathfrak{R}^{(2)P}$ as defined above is indeed bounded by the speed of light.

\providecommand{\href}[2]{#2}\begingroup\raggedright

\endgroup

\begin{thebibliography}{99}

\bibitem{maldacena}
J. M. Maldacena, \emph{The Large N Limit of Superconformal Field Theories and Supergravity}, \emph{Int.J.Theor.Phys.} {\bf 38} (1999) 1113--1133, [\href{http://xxx.lanl.gov/abs/hep-th/9711200}{hep-th/9711200}].

\bibitem{witten}
E. Witten, \emph{Anti-de Sitter space and holography}, \emph{Adv.Theor.Math.Phys.} {\bf 2} (1998) 253--291, [\href{http://xxx.lanl.gov/abs/hep-th/9802150}{hep-th/9802150}].

\bibitem{gubserklebanovpolyakov}
S. S. Gubser, I. R. Klebanov and A. M. Polyakov, \emph{Gauge Theory Correlators from Non-Critical String Theory}, \emph{Phys. Lett. B} {\bf 428} (1998) 105, [\href{http://xxx.lanl.gov/abs/hep-th/9802109}{hep-th/9802109}].

\bibitem{hubenyads}
V. E. Hubeny, \emph{The AdS/CFT Correspondence}, [\href{http://xxx.lanl.gov/abs/1501.00007}{arXiv:1501.00007}].

\bibitem{srednicki}
M. Srednicki, \emph{Entropy and Area}, \emph{Phys.Rev.Lett.} {\bf 71} (1993) 666--669, [\href{http://xxx.lanl.gov/abs/hep-th/9303048}{arXiv:hep-th/9303048}].

\bibitem{ecp}
J. Eisert, M. Cramer and M.B. Plenio, \emph{Area laws for the entanglement entropy - a review}, \emph{Rev. Mod. Phys.} {\bf 82} (2010) 277, [\href{http://xxx.lanl.gov/abs/0808.3773}{arXiv:0808.3773}].

\bibitem{calabresecardy2}
P. Calabrese and J. Cardy, \emph{Entanglement Entropy and Quantum Field Theory}, \emph{J.Stat.Mech.} {\bf 0406} (2004) P06002, [\href{http://xxx.lanl.gov/abs/hep-th/0405152}{arXiv:hep-th/0405152}].

\bibitem{casinihuertamyers}
H. Casini, M. Huerta, R. C. Myers, \emph{Towards a derivation of holographic entanglement entropy}, \emph{JHEP} {\bf 1105} (2011) 036, [\href{http://xxx.lanl.gov/abs/1102.0440}{arXiv:1102.0440}].

\bibitem{hubenyrangamanitakayanagi}
V. E. Hubeny, M. Rangamani and T. Takayanagi, \emph{A Covariant Holographic Entanglement Entropy Proposal}, \emph{JHEP} {\bf 0707} (2007) 062, [\href{http://xxx.lanl.gov/abs/0705.0016}{arXiv:0705.0016}].

\bibitem{hubenymaxfieldrangamanitonni}
V. E. Hubeny, H. Maxfield, M. Rangamani and E. Tonni, \emph{Holographic entanglement plateaux}, \emph{JHEP} {\bf 1308} (2013) 092, [\href{http://xxx.lanl.gov/abs/1306.4004}{arXiv:1306.4004}].

\bibitem{b-acs}
O. Ben-Ami, D. Carmi and J. Sonnenschein, \emph{Holographic Entanglement Entropy of Multiple Strips}, \emph{JHEP} {\bf 1411} (2014) 144, [\href{http://xxx.lanl.gov/abs/1409.6305}{arXiv:1409.6305}].

\bibitem{ryutakayanagi1}
S. Ryu and T. Takayanagi, \emph{Holographic Derivation of Entanglement Entropy from AdS/CFT}, \emph{Phys.Rev.Lett.} {\bf 96} (2006) 181602, [\href{http://xxx.lanl.gov/abs/hep-th/0603001}{arXiv:hep-th/0603001}].

\bibitem{ryutakayanagi2}
S. Ryu and T. Takayanagi, \emph{Aspects of Holographic Entanglement Entropy}, \emph{JHEP} {\bf 0608} (2006) 045, [\href{http://xxx.lanl.gov/abs/hep-th/0605073}{arXiv:hep-th/0605073}].

\bibitem{hartmanmaldacena}
T.Hartman and J. Maldacena, \emph{Time Evolution of Entanglement Entropy from Black Hole Interiors}, \emph{JHEP} {\bf 1305} (2013) 014, [\href{http://xxx.lanl.gov/abs/1303.1080}{arXiv:1303.1080}].

\bibitem{abajoapariciolopez}
J. Abajo-Arrastia, J. Aparicio, E. Lopez, \emph{Holographic Evolution of Entanglement Entropy}, \emph{JHEP} {\bf 1011} (2010) 149, [\href{http://xxx.lanl.gov/abs/1006.4090}{arXiv:1006.4090}].

\bibitem{bbbcckmsss}
V. Balasubramanian, A. Bernamonti, J. de Boer, N. Copland, B. Craps, E. Keski-Vakkuri, B. M\"{u}ller, A. Sch\"{a}fer, M. Shigemori and W. Staessens, \emph{Thermalization of Strongly Coupled Field Theories}, \emph{Phys.Rev.Lett.} {\bf 106} (2011) 191601, [\href{http://xxx.lanl.gov/abs/1012.4753}{arXiv:1012.4753}].

\bibitem{bbbcckmsss2}
V. Balasubramanian, A. Bernamonti, J. de Boer, N. Copland, B. Craps, E. Keski-Vakkuri, B. M\"{u}ller, A. Sch\"{a}fer, M. Shigemori and W. Staessens, \emph{Holographic Thermalization}, \emph{PhysRevD.} {\bf 84} (2011) 026010, [\href{http://xxx.lanl.gov/abs/1103.2683}{arXiv:1103.2683}].

\bibitem{bbccg}
V. Balasubramanian, A. Bernamonti, N. Copland, B. Craps and F. Galli, \emph{Thermalization of mutual and tripartite information in strongly coupled two dimensional conformal field theories}, \emph{PhysRevD.} {\bf 84} (2011) 105017, [\href{http://xxx.lanl.gov/abs/1110.0488}{arXiv:1110.0488}].

\bibitem{allaistonni}
A. Allais and E. Tonni, \emph{Holographic evolution of the mutual information}, \emph{JHEP} {\bf 1201} (2012) 102, [\href{http://xxx.lanl.gov/abs/1110.1607}{arXiv:1110.1607}].

\bibitem{lwwy}
Y.-Z. Li, S.-F. Wu, Y.-Q. Wang and G.-H. Yang, \emph{Linear growth of entanglement entropy in holographic thermalization captured by horizon interiors and mutual information}, \emph{JHEP} {\bf 1309} (2013) 057, [\href{http://xxx.lanl.gov/abs/1306.0210}{arXiv:1306.0210}].

\bibitem{liusuh1}
H. Liu, S. J. Suh, \emph{Entanglement Tsunami: Universal Scaling in Holographic Thermalization}, \emph{Phys. Rev. Lett.} {\bf 112} (2014) 011601 , [\href{http://xxx.lanl.gov/abs/1305.7244}{arXiv:1305.7244}].

\bibitem{liusuh2}
H. Liu, S. J. Suh, \emph{Entanglement growth during thermalization in holographic systems}, \emph{Phys. Rev. D} {\bf 89} (2014) 066012 , [\href{http://xxx.lanl.gov/abs/1311.1200}{arXiv:1311.1200}].

\bibitem{hubeny}
V. E. Hubeny, \emph{Extremal surfaces as bulk probes in AdS/CFT}, \emph{JHEP} {\bf 1207} (2012) 093, [\href{http://xxx.lanl.gov/abs/1203.1044}{arXiv:1203.1044}].

\bibitem{hubenymaxfield}
V. E. Hubeny and H. Maxfield, \emph{Holographic probes of collapsing black holes}, \emph{JHEP} {\bf 1403} (2014) 097, [\href{http://xxx.lanl.gov/abs/1312.6887}{arXiv:1312.6887}].

\bibitem{calabresecardy1}
P. Calabrese and J. Cardy, \emph{Evolution of Entanglement Entropy in One-Dimensional Systems}, \emph{J.Stat.Mech.} {\bf 0504} (2005) P04010, [\href{http://xxx.lanl.gov/abs/cond-mat/0503393}{arXiv:0503393}].

\bibitem{knstv}
V. Keranen, H. Nishimura, S. Stricker, O. Taanila, A. Vuorinen, \emph{Gravitational collapse of thin shells: Time evolution of the holographic entanglement entropy}, [\href{http://xxx.lanl.gov/abs/1502.01277}{arXiv:1502.01277}].

\bibitem{leichenauermoosa}
S. Leichenauer and M. Moosa, \emph{Entanglement Tsunami in (1+1)-Dimensions}, [\href{http://xxx.lanl.gov/abs/1505.04225}{arXiv:1505.04225}].

\bibitem{abgh}
C. T. Asplund, A. Bernamonti, F. Galli and T. Hartman, \emph{Entanglement Scrambling in 2d Conformal Field Theory}, [\href{http://xxx.lanl.gov/abs/1506.03772}{arXiv:1506.03772}].

\bibitem{hubenyrangamanitonnitonni}
V. E. Hubeny, M. Rangamani and E. Tonni, \emph{Thermalization of Causal Holographic Information}, \emph{JHEP} {\bf 1305} (2013) 136, [\href{http://xxx.lanl.gov/abs/1302.0853}{arXiv:1302.0853}].

\bibitem{amt}
M. Alishahiha, M. R. M. Mozaffar and M. R. Tanhayi, \emph{Evolution of Holographic n-partite Information}, [\href{http://xxx.lanl.gov/abs/1406.7677}{arXiv:1406.7677}].

\bibitem{hartman}
T. Hartman, \emph{Entanglement Entropy at Large Central Charge}, [\href{http://xxx.lanl.gov/abs/1303.6955}{arXiv:1303.6955}].

\bibitem{faulkner}
T. Faulkner, \emph{The Entanglement Renyi Entropies of Disjoint Intervals in AdS/CFT}, [\href{http://xxx.lanl.gov/abs/1303.7221}{arXiv:1303.7221}].

\end{thebibliography}
\end{document}